\documentclass[aip,cha,preprint,floats,amsmath,amssymb,superscriptaddress,showkeys]{revtex4-1}

\usepackage[utf8]{inputenc}
\usepackage[T1]{fontenc}
\usepackage{mathptmx}
\usepackage{amssymb}
\usepackage{amsmath} 

\usepackage{graphicx}
\usepackage{overpic}
\usepackage{multirow}
\usepackage{array}
\usepackage{tabularx, adjustbox} 
\usepackage{helvet}
\usepackage{booktabs}
\usepackage{algorithm}
\usepackage{algpseudocode}
\usepackage{mathrsfs}
\usepackage{hyperref}
\usepackage{xurl}
\usepackage{lineno}
\usepackage{graphicx}
\newsavebox{\grayleft}
\usepackage{xcolor}

\begin{document}

   


\title{Physics-guided curriculum learning for the identification of reaction-diffusion dynamics from partial observations}


\author{Hanyu Zhou}
\email{hanyuzhou@stu.pku.edu.cn}
\affiliation{HEDPS, Center for Applied Physics and Technology, and School of Mechanics and Engineering Science, Peking University, Beijing, 100871, China}

\author{Yuansheng Cao}
\email{yscao@tsinghua.edu.cn}
\affiliation{Department of Physics, Tsinghua University, Beijing, 100084, China}

\author{Yaomin Zhao}
\thanks{Corresponding author}
\email{yaomin.zhao@pku.edu.cn}
\affiliation{HEDPS, Center for Applied Physics and Technology, and School of Mechanics and Engineering Science, Peking University, Beijing, 100871, China}



\begin{abstract}
Reaction-diffusion (RD) systems provide fundamental models for understanding self-organized spatiotemporal patterns across natural and engineered settings, yet reliable parameter estimation remains challenging, particularly when observations are sparse, noisy, and restricted to a subset of state variables.
We introduce CLIP (Curriculum Learning Identification via PINNs), a physics-guided framework built on physics-informed neural networks for joint parameter inference and hidden-state reconstruction under partial observability.
Leveraging the locally reaction-dominated dynamics for initialization, the CLIP training progresses from reaction-dominated regimes to full spatiotemporal dynamics
using curriculum learning and an anchored widening transfer strategy.
Across three canonical RD benchmarks, CLIP achieves more accurate and robust identification than baseline methods.
Furthermore, the CLIP framework is successfully applied to infer the dynamics of the Min system in bacteria, where only membrane-bound species are observed and key kinetic rates span multiple orders of magnitude.
Ablation experiments and loss-landscape visualizations demonstrate that both the curriculum stages and the anchored transfer are essential for stable convergence.

\end{abstract}




\maketitle

\begin{quotation}
Reaction-diffusion systems describe how local reactions and diffusion generate spatial patterns in chemical and biological systems. 
Experimental measurements often capture only a subset of the interacting components, so physical parameters and unobserved states must be inferred together. 
Errors in these estimates can compensate for one another, producing a close fit to the observations even when the inferred dynamics are inaccurate. 
Here, we use locally reaction-dominated dynamics to guide parameter identification and hidden-state reconstruction from partial observations. 
Numerical tests on three canonical reaction-diffusion systems and a model of bacterial Min-protein organization show more accurate parameter estimates and better recovery of the unobserved fields than a standard physics-informed neural network.
\end{quotation}


\section{Introduction}
\label{sec:intro}

Many dynamical systems in physics, engineering, and biology are only partially observable in experiments \cite{daniele2013flame,loose2008minde,rathinam2021chemical}.
Measurements often cover only a subset of state variables, while the unobserved components still interact nonlinearly with the measured ones. 
Under such partial state access, jointly recovering governing parameters and hidden states faces practical identifiability challenges, as weak observational constraints allow errors in parameter estimates and state reconstructions to compensate for one another.
A representative example is the bacterial Min system \cite{loose2008minde}, in which membrane-associated species can be monitored by fluorescence imaging, whereas cytosolic concentrations are difficult to quantify.
These challenges have motivated a growing body of work on parameter identification and state reconstruction under partial state observations.


\subsection{Related work}
Recent advances in machine learning have motivated extensive efforts to model and identify dynamical systems from partial state observations.
One direction formulates the problem as spatiotemporal reconstruction or forecasting from sparse measurements. 
Recent generative approaches use pretrained score-based or diffusion models \cite{song2021scorebased} to reconstruct spatiotemporal fields from sparse observations \cite{Huang2024DiffusionPDE,li2024S3GM}, and later work further introduces physically consistent sampling for forward and inverse problems \cite{Jacobsen2025CoCoGen}. 
Delay-based architectures \cite{churchill2023dnn,churchill2025dnn} have also been proposed to predict the evolution of systems governed by partial differential equations (PDEs) from incomplete observations by exploiting histories of the observable variables.
However, these methods are designed primarily for state reconstruction or prediction rather than for explicit identification of the governing equations and physical parameters. 
Consequently, their physical interpretability remains limited. 
Without explicit identification of the governing parameters, they cannot be readily used to predict system behavior under novel operating conditions, which is essential in applications such as the design of pattern-forming chemical reactors and the analysis of biological oscillators.

An alternative direction aims to explicitly recover the underlying governing dynamics by incorporating physical knowledge.
A foundational result in this direction is Takens' embedding theorem \cite{takens2006takenstheory}, which guarantees that, under generic conditions, delay-coordinate maps constructed from a single observable can embed the attractor of an autonomous dynamical system. 
Building on this principle, partially observed ordinary differential equation (ODE) systems have been identified using delay embeddings \cite{wu2024low-dimensional} or by learning diffeomorphic latent representations \cite{bakarji2023diffeo}.
Complementary approaches adopt a model-selection perspective, screening candidate model spaces via sensitivity analysis and clustering \cite{stepaniants2024HDI} or performing sparse regression over predefined function libraries with variational annealing \cite{ribera2022dahsi}. 
While these methods successfully recover governing laws from partial measurements, their demonstrated applicability is largely restricted to ODE systems that admit low-dimensional state-space representations. 
Extension to spatially extended PDE systems, in which the state dimension scales with spatial discretization and diffusion introduces spatial coupling, remains an open challenge.

Physics-informed neural networks (PINNs) \cite{raissi2019pinn} provide a flexible framework for inverse PDE problems by incorporating observed data and differential equation residuals into a single training objective.
Nonetheless, PINNs can converge slowly or fail to learn accurate solutions because of imbalanced training dynamics and ill-conditioning \cite{wang2022pinnkernel,Krishnapriyan2021pinnfail,cao2025analysis}.
These difficulties can be compounded when governing parameters and hidden states must be inferred jointly from partial state observations.
Weak observational constraints can allow changes in parameter estimates to be offset by changes in inferred hidden states, yielding similar fits to the measurements.
Existing methods incorporate dynamical constraints into inference from partial observations, using flexibly placed residual points to enforce governing equations \cite{raissi2020hiddenfluid} or matching higher-order time derivatives to constrain the inferred dynamics \cite{lu2022CNN}.
These difficulties can be further aggravated in partially observed systems with stiff and multiscale dynamics, where parameter sensitivities may vary across the domain and the network needs to resolve widely separated spatial and temporal scales.
These challenges motivate training strategies that exploit the physical structure of the governing equations, with the aim of decomposing the inverse problem into tractable subproblems, thereby alleviating the optimization difficulties in standard PINNs.


\subsection{Main Contributions}
Reaction-diffusion (RD) systems are widely used due to their success in explaining pattern formation across chemistry \cite{turing1990chemical}, biology \cite{howard2011turing,kondo2010ApplyBiological,collinet2021ApplyMorphogenesis}, ecology \cite{holmes1994ecology}, and materials science \cite{tan2018ApplyPolyamide,gu2023Applynanotwins}.
Accurately identifying RD parameters remains challenging, particularly under partial observations. 
This difficulty arises partly because RD patterns are often governed by stiff and nonlinear dynamics, where steep gradients and fine-scale features render the system highly sensitive to parameter variations.
This sensitivity intensifies under partial and noisy observations, particularly when the patterns contain sharp fronts and fine-scale structures, leading to a rugged and non-convex optimization landscape.
Consequently, obtaining robust parameter estimates for RD systems remains a significant and unresolved challenge.

\begin{figure}[!htbp]%
\centering
\includegraphics[width=1.0\textwidth]{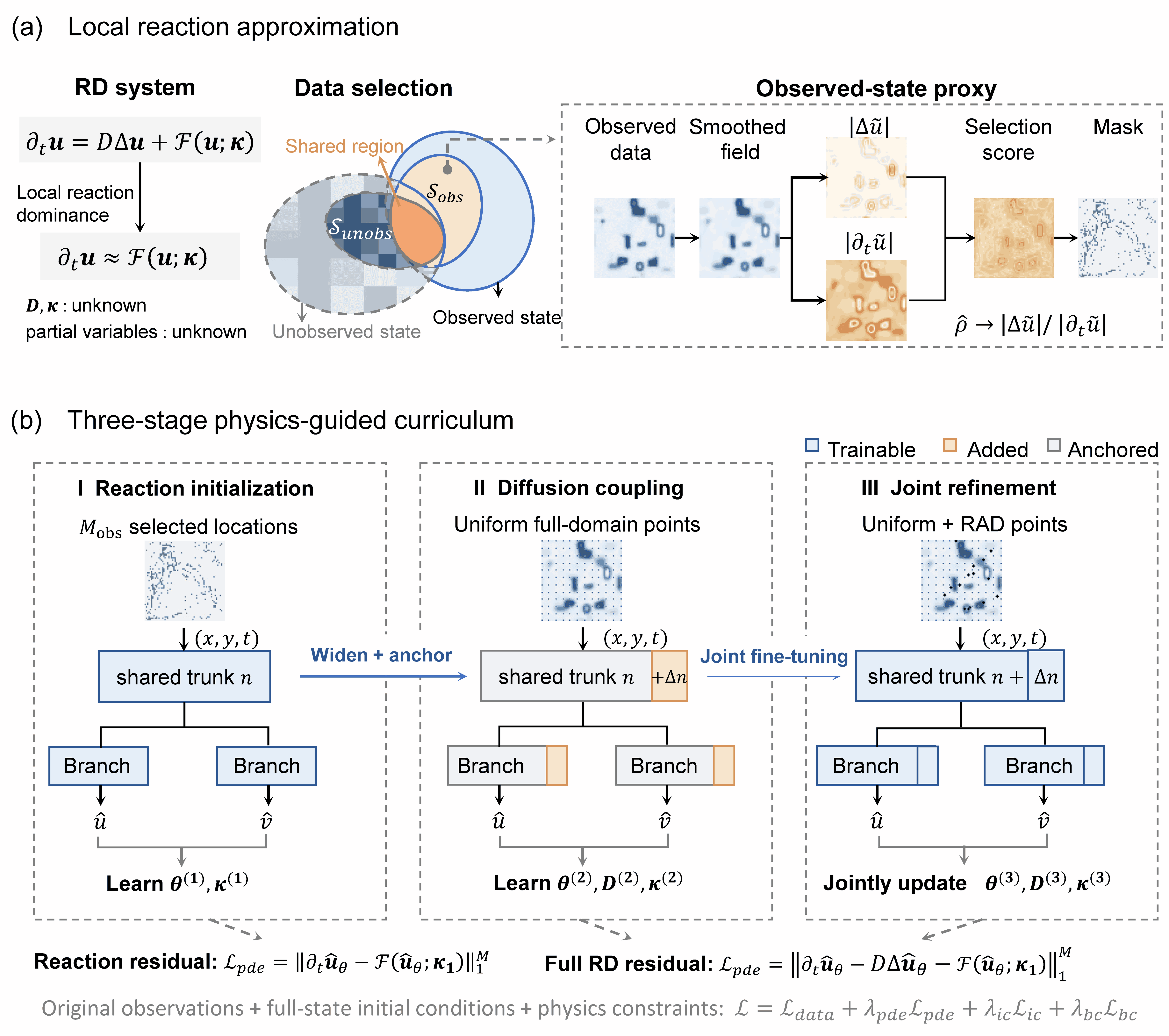}
\caption{\textbf{The proposed CLIP framework.}
\textbf{(a)} Local reaction approximation and observed-state proxy.
The intersection of component-wise reaction-dominated regions (orange) defines a shared region admitting a local coupled-ODE approximation.
With hidden states and physical coefficients unknown, a proxy based on the Laplacian and temporal derivative magnitudes of the smoothed observed field is used to construct the mask $M_{\mathrm{obs}}$ (Section~\ref{subsec:stage1}).
Field thumbnails illustrate the Lotka--Volterra example.
\textbf{(b)} Three-stage physics-guided curriculum.
A shared network trunk maps the coordinate $(\mathbf{x},t)$ to task-specific branches for the observed and unobserved state variables.
Stage~I initializes $\theta$ and $\kappa$ using mask-selected data and reaction residuals.
Stage~II widens the network from $n$ to $n+\Delta n$ and introduces full reaction--diffusion constraints at uniform points, with reduced-learning-rate updates for inherited weights and $\kappa$.
Stage~III removes anchoring and freezing restrictions and jointly refines $\theta$, $\kappa$, and $D$ using uniform and residual-adaptive (RAD) points.
Blue, orange, and gray denote trainable, added, and anchored network components, respectively.
Superscript $M$ indicates mask-selected loss terms.
}
\label{fig:workflow}
\end{figure}

To alleviate these optimization difficulties, we introduce physics-guided Curriculum Learning Identification via PINNs (CLIP), a framework that couples curriculum learning \cite{bengio2009curriculum,wang2021curriculumsurvey} with PINNs for parameter identification and hidden-state reconstruction in partially observed RD systems.
The core idea of CLIP is to use locally reaction-dominated dynamics to initialize joint parameter and state inference, then progressively incorporate diffusion through a three-stage curriculum (Fig.~\ref{fig:workflow})
Specifically, a proxy constructed from the observed variables in Fig.~\ref{fig:workflow}a is used to select candidate regions, where the reaction parameters and state representation are initialized through an approximate local ODE system.
The resulting estimates initialize joint inference of the state fields, reaction parameters, and diffusion coefficients under the full PDE constraints.
To stabilize this transition, an anchored widening transfer strategy is employed to expand network capacity while updating inherited weights and reaction parameters at a reduced learning rate.
In the final stage, anchoring is removed and all trainable parameters are jointly refined under the full PDE constraints, while residual-adaptive sampling focuses on regions with larger physics residuals.
CLIP is validated on three benchmark RD systems and a biologically motivated Min model using partial observations and initial profiles of all state variables.
Baseline comparisons and ablation studies show the benefits of the curriculum and anchored widening transfer for parameter identification and hidden-state reconstruction under the tested noise conditions.

The paper is organized as follows. Section 2 introduces the preliminaries, including the inverse problem formulation and the PINN-based framework for PDE identification under partial observations.
The proposed CLIP framework is then presented in detail in Section 3.
Section 4 presents the benchmark problems and numerical results, including baseline comparisons and ablation studies. 
Section 5 discusses the applicability conditions, limitations, and future extensions of the framework.

\section{Preliminaries}
\subsection{Problem formulation}\label{sec:problem}
In general, an RD system is described by partial differential equations of the form
\begin{equation}
\partial_t \mathbf{u}
= \boldsymbol{D} \Delta\mathbf{u}
+ \mathcal{F}\!\left(\mathbf{u};\boldsymbol{\kappa}\right),
\label{eq:inv}
\end{equation}
where $\mathbf{u}:\Omega\times(0,T]\to\mathbb{R}^N$ denotes the vector of state variables, 
$\mathbf{x} \in \Omega \subset \mathbb{R}^d$ is the spatial coordinate, $t \in [0,T]$ is the time variable.
Here $\Delta = \sum_{i=1}^d \partial^2/\partial x_i^2$ denotes the Laplacian. $\boldsymbol{D}=\mathrm{diag}(D_1,\ldots,D_N)$ collects the diffusion coefficients, and $\boldsymbol{\kappa}$ denotes the unknown reaction parameters in $\mathcal{F}$.
Under partial observability, only a subset of the state variables is accessible through noisy measurements, while the remaining components are unobserved.
To formalize the observation process, we introduce an observation operator $\mathcal{H}:\mathbb{R}^N \to \mathbb{R}^{N_m}$ with ${N_m} \le N$ that maps the full state to the measurable components. The observed data are written as
\begin{equation}
\mathbf{y}(\mathbf{x},t) = \mathcal{H}\!\left(\mathbf{u}(\mathbf{x},t)\right) + \boldsymbol{\varepsilon}(\mathbf{x},t),
\end{equation}
where $\mathbf{y}$ denotes the measurements and $\boldsymbol{\varepsilon}$ represents observational noise arising from experimental acquisition and preprocessing.
In addition to the partial observations for $t>0$, we assume that
the initial spatial profiles of all state variables are known,
\begin{equation}
\mathbf{u}(\mathbf{x},0)=\mathbf{u}_0(\mathbf{x}),
\qquad \mathbf{x}\in\Omega.
\label{eq:initial_state}
\end{equation}
These initial profiles provide key constraints for reconstructing
unobserved states and can be established through controlled
preparation or calibration.
Given the partial observations $\mathbf{y}$ and the full initial
state $\mathbf{u}_0$, the inverse problem consists of identifying
the reaction parameters $\boldsymbol{\kappa}$ and diffusion
coefficients $\boldsymbol{D}$, and reconstructing the unobserved
state components over $\Omega\times(0,T]$.



\subsection{PINN formulation of the inverse problem}\label{subsec:CLIPmethod}

We formulate state reconstruction and parameter estimation as a
joint optimization problem using physics-informed neural networks.
Let $\hat{\mathbf{u}}(\mathbf{x},t;\boldsymbol{\theta})$ denote
the neural network approximation of the full state, where
$\boldsymbol{\theta}$ collects the trainable network parameters.
The reaction parameters $\boldsymbol{\kappa}$ and diffusion
coefficients $\boldsymbol{D}$ are optimized together with
$\boldsymbol{\theta}$.
The physics residual of the inverse problem is defined as
\begin{equation}
\mathcal{R}(\mathbf{x},t;\boldsymbol{\theta},\boldsymbol{\kappa},\boldsymbol{D})
=
\partial_t \hat{\mathbf{u}}
-
\boldsymbol{D}\Delta \hat{\mathbf{u}}
-
\mathcal{F}\!\left(\hat{\mathbf{u}};\boldsymbol{\kappa}\right),
\end{equation}
where all derivatives are computed through automatic differentiation.

Under partial observability, only measurements of $\mathcal{H}(\mathbf{u})$ are available. Given observation data $\{(\mathbf{x}_i,t_i),\mathbf{y}_i\}_{i=1}^{N_d}$, a generic data discrepancy term is written as
\begin{equation}\label{eq:dataloss}
\mathcal{L}_{\mathrm{data}}
=
\frac{1}{N_d}
\sum_{i=1}^{N_d}
\left\|
\mathcal{H}\!\left(\hat{\mathbf{u}}(\mathbf{x}_i,t_i;\boldsymbol{\theta})\right)
-
\mathbf{y}_i
\right\|.
\end{equation}
To enforce the governing dynamics, a set of collocation points $\{(\mathbf{x}_r,t_r)\}_{r=1}^{N_r}$ is sampled in the spatiotemporal domain, leading to the residual penalty
\begin{equation}\label{eq:pdeloss}
\mathcal{L}_{\mathrm{pde}}
=
\frac{1}{N_r}
\sum_{r=1}^{N_r}
\left\|
\mathcal{R}(\mathbf{x}_r,t_r;\boldsymbol{\theta},\boldsymbol{\kappa},\boldsymbol{D})
\right\|.
\end{equation}
The prescribed initial profiles are enforced through
\begin{equation}\label{eq:icloss}
\mathcal{L}_{\mathrm{ic}}
=
\frac{1}{N_{ic}}
\sum_{j=1}^{N_{ic}}
\left\|
\hat{\mathbf{u}}(\mathbf{x}_j,0;\boldsymbol{\theta})
-
\mathbf{u}_0(\mathbf{x}_j)
\right\|.
\end{equation}
Periodic boundary conditions are enforced by matching all state
components and their first spatial derivatives
\begin{equation}\label{eq:bcloss}
\begin{aligned}
\mathcal{L}_{\mathrm{bc}}
=
\frac{1}{N_{\mathrm{bc}}}
\sum_{k=1}^{N_{\mathrm{bc}}}
\Big[
&\ell\!\left(
\hat{\mathbf{u}}(\mathbf{x}_k,t_k)
-
\hat{\mathbf{u}}(\mathbf{x}'_k,t_k)
\right)
\\
&+
\ell\!\left(
\partial_n\hat{\mathbf{u}}(\mathbf{x}_k,t_k)
-
\partial_n\hat{\mathbf{u}}(\mathbf{x}'_k,t_k)
\right)
\Big],
\end{aligned}
\end{equation}
where $N_{\mathrm{bc}}$ is the total number of sampled boundary
pairs, and $\partial_n$ denotes differentiation along the same coordinate direction normal to each boundary pair.

Under noisy or partially observed settings, the norms in Eqs.~\eqref{eq:dataloss}-\eqref{eq:icloss} are implemented using the Smooth $L_1$ loss for improved robustness \cite{girshick2015fast}.
The network parameters and physical coefficients are jointly
estimated by minimizing
\begin{equation}\label{eq:joint_optimization}
\min_{\boldsymbol{\theta},\boldsymbol{\kappa},\boldsymbol{D}}
\mathcal{L}
=
\mathcal{L}_{\mathrm{data}}
+
\lambda_{\mathrm{pde}}\mathcal{L}_{\mathrm{pde}}
+
\lambda_{\mathrm{ic}}\mathcal{L}_{\mathrm{ic}}
+
\lambda_{\mathrm{bc}}\mathcal{L}_{\mathrm{bc}},
\end{equation}
where the weights control the relative contributions of the
physics, initial-condition, and boundary-condition losses.
In partially observed systems, joint estimation introduces an additional challenge to PINN optimization \cite{wang2021PINNgrad,Krishnapriyan2021pinnfail}.
Unobserved state trajectories are constrained by the governing
equations and the initial and boundary conditions, while the
physical parameters in these equations also need to be estimated.
This coupling allows state and parameter errors to partially
offset each other during training, complicating joint optimization.
This motivates a physics-guided curriculum that organizes
joint estimation into a sequence of more tractable
optimization subproblems.


\section{A Physics-guided curriculum for parameter and state estimation}\label{sec:clip}
We address this difficulty by exploiting a common feature of pattern organization in reaction--diffusion systems.
During the formation and evolution of spatiotemporal patterns, such as propagating waves, reaction kinetics and diffusive transport act jointly but contribute differently across local regions.
We use regions where reaction dominates to initialize the reaction parameters and state fields.
Building on this observation, we propose CLIP, a three-stage physics-guided curriculum that progresses from reaction-based initialization to joint refinement under the full reaction--diffusion constraints.

\subsection{Primary stage \uppercase\expandafter{\romannumeral1}: Reaction-dominated initialization}\label{subsec:stage1}
The primary stage draws conceptual motivation from Takens' embedding theorem \cite{takens2006takenstheory}, which shows that, under suitable conditions, delay coordinates constructed from a single observable can embed the state space of a finite-dimensional autonomous system.
Although this theorem does not directly apply to  RD systems, whose continuous spatial state is infinite-dimensional, it motivates considering regions where diffusion contributions are small relative to the corresponding reaction terms. 
In such regions, the local dynamics may be approximated by a finite-dimensional reaction ODE, in which temporal observations of a single variable may carry information about the coupled reaction dynamics.
This observation suggests that, when such regions exist and can be identified through a physically meaningful criterion, they provide a suitable subset of samples for initializing the reaction parameters and state representation.


To characterize these regions, we examine the pointwise reaction--diffusion balance of all $N$ state variables.
For a solution of the full reaction-diffusion equations, the reaction residual for each component satisfies
\begin{equation}
r_i^{\mathrm{R}}(\mathbf{x},t)
:=
\partial_t u_i
-
\mathcal{F}^i(\mathbf{u};\boldsymbol{\kappa})
=
D_i\Delta u_i,
\qquad i=1,\ldots,N.
\label{eq:reaction_residual_balance}
\end{equation}
Since this residual equals the omitted diffusion contribution, we quantify the local reaction-approximation error through the regularized ratio
\begin{equation}
e_i(\mathbf{x},t)
=
\frac{|r_i^{\mathrm{R}}(\mathbf{x},t)|}
{|\mathcal{F}^i(\mathbf{u};\boldsymbol{\kappa})|+\eta_i}
=
\frac{|D_i\Delta u_i|}
{|\mathcal{F}^i(\mathbf{u};\boldsymbol{\kappa})|+\eta_i},
\label{eq:reaction_approximation_indicator}
\end{equation}
where $\eta_i>0$ is a small regularization scale.
Since the primary stage uses the local reaction approximation for all state variables, we define the common reaction-dominated set as
\begin{equation}
\mathcal{S}_{\varepsilon}
=
\bigcap_{i=1}^{N}
\left\{
(\mathbf{x},t):
e_i(\mathbf{x},t)\leq\varepsilon
\right\}.
\label{eq:common_reaction_dominated_set}
\end{equation}
Here $\varepsilon>0$ is a prescribed small tolerance.
This common condition provides a basis for initialization using the coupled reaction residuals.

However, $\mathcal{S}_{\varepsilon}$ depends on the full state and unknown physical parameters and therefore cannot be determined directly from the available partial observations.
When reaction coupling produces shared regions with weak spatial variation and appreciable reaction activity across components, the observed variables may provide useful information for locating samples in the common reaction-dominated set.
We therefore use the observed components to select candidate regions for Stage I.
For each observed component $i=1,\ldots,N_m$, where $N_m<N$, we use the ratio of the Laplacian magnitude to the temporal derivative magnitude as a score.
Since $D_i>0$ is constant, this score has the same ordering as the magnitude of the ommited diffusion term relative to $|\partial_t u_i|$ wherever $\partial_t u_i \neq 0$. We write
\begin{equation}
\begin{aligned}
\rho_i(\mathbf{x},t)
&=
\frac{|\Delta u_i|}{|\partial_t u_i|},
\end{aligned}
\label{eq:observable_reaction_proxy}
\end{equation}
and estimate the derivatives from the observed field.

For noisy observations, we first fit an auxiliary smoother to the observed variables and compute the temporal derivatives and Laplacians of the resulting smoothed fields $\widetilde{u}_i$.
The practical proxy is evaluated as
\begin{equation}
\widehat{\rho}_i(\mathbf{x},t)
=
\frac{|\Delta\widetilde{u}_i(\mathbf{x},t)|+c_{\Delta,i}}
{|\partial_t\widetilde{u}_i(\mathbf{x},t)|+c_{t,i}},
\label{eq:smoothed_reaction_proxy}
\end{equation}
where
$c_{\Delta,i}=\epsilon_{\mathrm{num}}\langle|\Delta\widetilde{u}_i|\rangle$
and
$c_{t,i}=\epsilon_{\mathrm{num}}\langle|\partial_t\widetilde{u}_i|\rangle$,
with $\epsilon_{\mathrm{num}}=10^{-8}$ and $\langle\cdot\rangle$ denoting the average over the candidate spatiotemporal points.
These numerical floors scale with the corresponding derivative magnitudes and regularize the ratio when the estimated derivatives approach zero.
The observation-based reaction-dominance mask is then defined as
\begin{equation}
M_{\mathrm{obs}}(\mathbf{x},t)
=
\prod_{i=1}^{N_m}
\mathbb{I}\!\left(
\widehat{\rho}_i(\mathbf{x},t)
<
\tau_{\rho}q_{i,99}
\right),
\label{eq:observed_reaction_mask}
\end{equation}
where $\mathbb{I}(\cdot)$ is the indicator function and $q_{i,99}$ is the 99th percentile of $\widehat{\rho}_i$ over the candidate points.
This percentile provides a componentwise reference scale that reduces sensitivity to extreme ratio values.
We set $\tau_{\rho}=10^{-3}$ and require all observed components to satisfy the selection criterion.

During the primary stage, the objective combines the data loss evaluated against the original observations at locations selected by $M_{\mathrm{obs}}$, the reaction-residual loss for all components at collocation points selected by the same mask, and the initial-condition loss for all state variables.
The smoother is used only for mask construction, while the data loss is evaluated against the original observations at the selected locations and the reaction residuals are computed from the main network outputs through automatic differentiation.
The network parameters $\theta$ and reaction parameters $\boldsymbol{\kappa}$ are optimized jointly, while the diffusion coefficients are not optimized at this stage.
The resulting state representation and reaction parameters, denoted by $(\theta^{(1)},\boldsymbol{\kappa}^{(1)})$, provide the initialization for subsequent training with the full reaction--diffusion equations.

\subsection{Middle stage \uppercase\expandafter{\romannumeral2}: Diffusive coupling via anchored widening transfer}
Once the network and reaction parameters have been initialized in the primary stage, we proceed to joint state and parameter estimation under the full PDE constraints.
This transition incorporates second-order diffusion terms with trainable diffusion coefficients into the physics residual and extends training from selected local samples to uniformly sampled points across the full spatiotemporal domain.
However, naive fine-tuning under this new objective may substantially alter the initial reaction estimates and the learned state representation.
The added spatial-derivative constraints change the optimization gradients, and the optimizer may initially reduce the residual through compensatory changes in the reaction parameters before the state representation has adapted to these constraints.
To stabilize this transition while expanding the model's capacity to represent diffusive coupling, we introduce an anchored widening transfer strategy \cite{rusu2016progressive,xuhong2018L2-SP}.

As illustrated in Fig.~\ref{fig:workflow}(c), each hidden layer of the pretrained network is widened by adding a small number of neurons, providing additional trainable degrees of freedom to fit the full spatiotemporal field and satisfy the added physics constraints.
For a pretrained network with weight matrices $W^{(l)} \in \mathbb{R}^{n_{l+1}\times n_l}$, the widened matrices take the block form
\begin{equation}
\tilde{W}^{(l)}
=
\begin{bmatrix}
W_{oo}^{(l)} & W_{on}^{(l)} \\
W_{no}^{(l)} & W_{nn}^{(l)}
\end{bmatrix},
\end{equation}
where subscripts $o$ and $n$ index inherited and newly introduced units, respectively.
Here $W_{oo}^{(l)}$ maps inherited inputs to inherited outputs,
$W_{on}^{(l)}$ maps new inputs to inherited outputs,
$W_{no}^{(l)}$ maps inherited inputs to new outputs,
and $W_{nn}^{(l)}$ maps new inputs to new outputs.
The block $W_{oo}^{(l)}$ inherits the weights learned in the primary stage, while the newly introduced blocks use Xavier initialization.
The inherited blocks and reaction coefficients are updated with a reduced learning rate $\eta_a<\eta$, while the remaining trainable weights and diffusion coefficients use the default rate $\eta$.
Detailed architecture, parameter grouping, and optimization settings are provided in Appendix~\ref{SI:main_architecture}.
By the end of the middle stage, the model provides preliminary joint estimates of the state fields and the reaction and diffusion coefficients under the full PDE constraints.
The subsequent stage removes anchoring and uses residual-based adaptive distribution (RAD) sampling to add collocation points in regions of high physics residual for further joint refinement.

\subsection{Fine-tune stage \uppercase\expandafter{\romannumeral3}: Global refinement with residual-adaptive sampling}
After primary and middle stages, the reaction parameters and the broad spatial structure of
the diffusive coupling have been established, but residual errors typically concentrate in narrow
regions of intense spatiotemporal activity for instance, near reaction fronts, pulse peaks, or
transient interface regions. Uniform collocation sampling under-resolves these regions and can
leave a small but systematic bias in the identified coefficients. The third stage therefore couples
joint refinement of all parameters with RAD sampling, which redistributes collocation points
toward regions of high physics residual.

The fine-tuning stage removes the anchoring and freezing restrictions and jointly updates all trainable parameters under the full PDE objective using a unified learning rate.
Let $\hat{\mathbf{u}}_\theta(\mathbf{x},t)=\big(\hat{u}_\theta^{1},\ldots,\hat{u}_\theta^{N}\big)$ denote the network prediction of the $N$ state variables, where $\theta$ collects the neural network parameters.
Given candidate spatiotemporal points $\{(\mathbf{x}_k,t_k)\}_{k=1}^{N_k}$, the residual magnitude for each variable is defined as
\begin{equation}
\varepsilon_k^{i}
=
\left|
\partial_t \hat{u}_\theta^{i}(\mathbf{x}_k,t_k)
-
D_i \Delta \hat{u}_\theta^i(\mathbf{x}_k,t_k)
-
\mathcal{F}^{i}\big(\hat{\mathbf{u}}_\theta(\mathbf{x}_k,t_k);\boldsymbol{\kappa}\big)
\right|.
\end{equation}
To reduce the influence of differences in residual scale across variables, residual magnitudes are normalized by
\begin{equation}
\tilde{\varepsilon}_k^{i}
=
\frac{\varepsilon_k^{i}}{\max_{1\le j\le N_k}\varepsilon_j^{i}},
\end{equation}
and averaged across variables to obtain a pointwise score
\begin{equation}
\bar{\varepsilon}_k=\frac{1}{N}\sum_{i=1}^{N}\tilde{\varepsilon}_k^{i}.
\end{equation}
The discrete sampling distribution is defined as
\begin{equation}
p_k=\frac{\bar{\varepsilon}_k}{\sum_{j=1}^{N_k}\bar{\varepsilon}_j}.
\end{equation}
Additional collocation points are sampled according to $\{p_k\}_{k=1}^{N_k}$ and used for joint parameter and state refinement, as illustrated in the fine-tuning stage of Fig.~\ref{fig:workflow}(b).

\section{Numerical experiments}
The CLIP framework is evaluated on three canonical reaction-diffusion systems, including the $\lambda$-$\omega$ model, the Gray-Scott system and the Lotka-Volterra system.
To assess its performance in a biological setting, we then consider a model of Min-protein self-organization.
These cases are designed to assess parameter identification and hidden-state reconstruction under partial and noisy observations across systems with distinct reaction kinetics, spatial patterns, and diffusion scales.
For all systems, reference solutions are generated using a spectral solver with periodic boundary conditions. 
The computational domains and spatiotemporal resolutions are summarized in Table~\ref{tab:dataset}.
Training data are obtained by
uniformly sampling 2\% of the full spatiotemporal grid from the observable variables.
\begin{table}[!htbp]
\caption{Computational domains and spatiotemporal grids of the datasets.}
\label{tab:dataset}
\small
\setlength{\tabcolsep}{8pt}
\begin{tabular*}{\textwidth}{@{\extracolsep{\fill}}lcc@{}}
\toprule
PDE system & Domain $(x,y,t)$ & Grid $(N_x,N_y,N_t)$ \\
\midrule
$\lambda$-$\omega$ RD & $[0,20]^2 \times [0,100]$ & $(64,64,1000)$ \\
Gray-Scott & $[0,100\sqrt{5}]^2 \times [0,500]$ & $(64,64,500)$ \\
Lotka-Volterra & $[0,64]^2 \times [0,100]$ & $(64,64,1000)$ \\
Min system & $[0,900]^2 \times [0,2000]$ & $(120,120,2000)$ \\
\bottomrule
\end{tabular*}
\end{table}

To assess robustness, additive Gaussian noise is applied to the observable variables as
\begin{equation}
\mathbf{u}_{\mathrm{obs}}(\mathbf{x},t)
=
\mathbf{u}(\mathbf{x},t)
+
\sigma \cdot \mathrm{std}\big(\mathbf{u}(\mathbf{x},t)\big)\cdot \mathscr{N}(0,1),
\end{equation}
where $\sigma$ denotes the noise level, set to $0$ for clean observations and $5\%$ or $10\%$ for noisy observations.
Network architectures, optimization settings, and further implementation details are provided in~\ref{SI:hyperparameters}.
To assess its performance, CLIP is compared with a standard PINN baseline, the ensemble Kalman filter (EnKF) \cite{strofer2021dafi}, and particle swarm optimization (PSO) \cite{wang2018pso}.
Ablation studies are conducted by progressively adding curriculum learning and the anchored widening strategy to the baseline PINN.

Parameter identification is evaluated using the relative absolute error (RAE) of each coefficient,
\begin{equation}
\mathrm{RAE}_{p}
=
\frac{\lvert \nu_p^{\mathrm{pred}} - \nu_p^\ast \rvert}
{\lvert \nu_p^\ast \rvert},
\end{equation}
where $\nu_p^{\mathrm{pred}}$ and $\nu_p^\ast$ denote the estimated and reference values of the $p$th parameter, respectively.
The mean relative absolute error (MRAE) over all $P$ identified parameters is defined as
\begin{equation}
\mathrm{MRAE}
=
\frac{1}{P}\sum_{p=1}^{P}\mathrm{RAE}_{p},
\end{equation}
where $P$ is the total number of identified parameters.
Hidden-state reconstruction is quantified by the relative $L_2$ error,
\begin{equation}
\mathrm{RL2E}_i
=
\frac{\left\|\hat{u}_i-u_i^\ast\right\|}
{\left\|u_i^\ast\right\|} ,
\end{equation}
where $\hat{u}_i$ and $u_i^\ast$ denote the reconstructed and reference hidden fields, respectively. 

\subsection{Case 1: The $\lambda$-$\omega$ reaction-diffusion system}
The $\lambda$-$\omega$ reaction-diffusion system is considered as a canonical benchmark for validating the recovery of hidden dynamics under partial observations.
This system exhibits smooth oscillatory behavior, providing a suitable testbed for proof-of-concept verification.
We study the two-dimensional system with periodic boundary conditions,
\begin{equation}
u_t = D_1 \Delta u + \lambda(A)u - \omega(A)v,
\end{equation}
\begin{equation}
v_t = D_2 \Delta v + \omega(A)u + \lambda(A)v,
\end{equation}
where $u$ and $v$ denote the state variables, $A=u^2+v^2$,
$\lambda(A)=\alpha(1-A)$, and $\omega(A)=\beta A$.
In this experiment, only the field $u$ is observed.
The unknown coefficients $\alpha$, $\beta$, $D_1$, and $D_2$ are inferred while the unobserved field $v$ is simultaneously reconstructed.
\begin{figure}[!htbp]
    \centering
    \begin{minipage}[t]{0.505\textwidth}
        \vspace{0pt}
        \includegraphics[width=\linewidth]
        {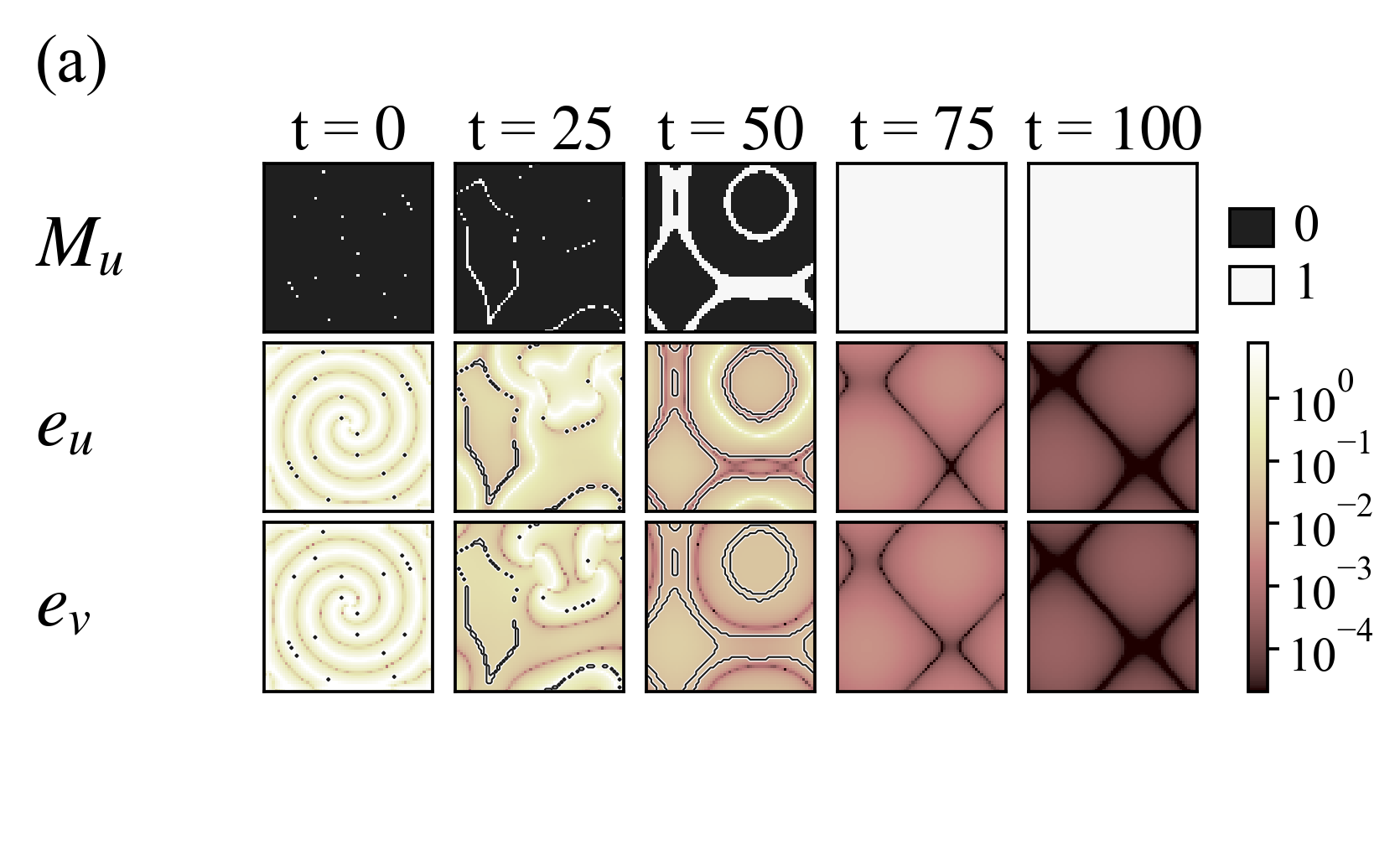}\par
        \nointerlineskip
        \includegraphics[width=\linewidth]
        {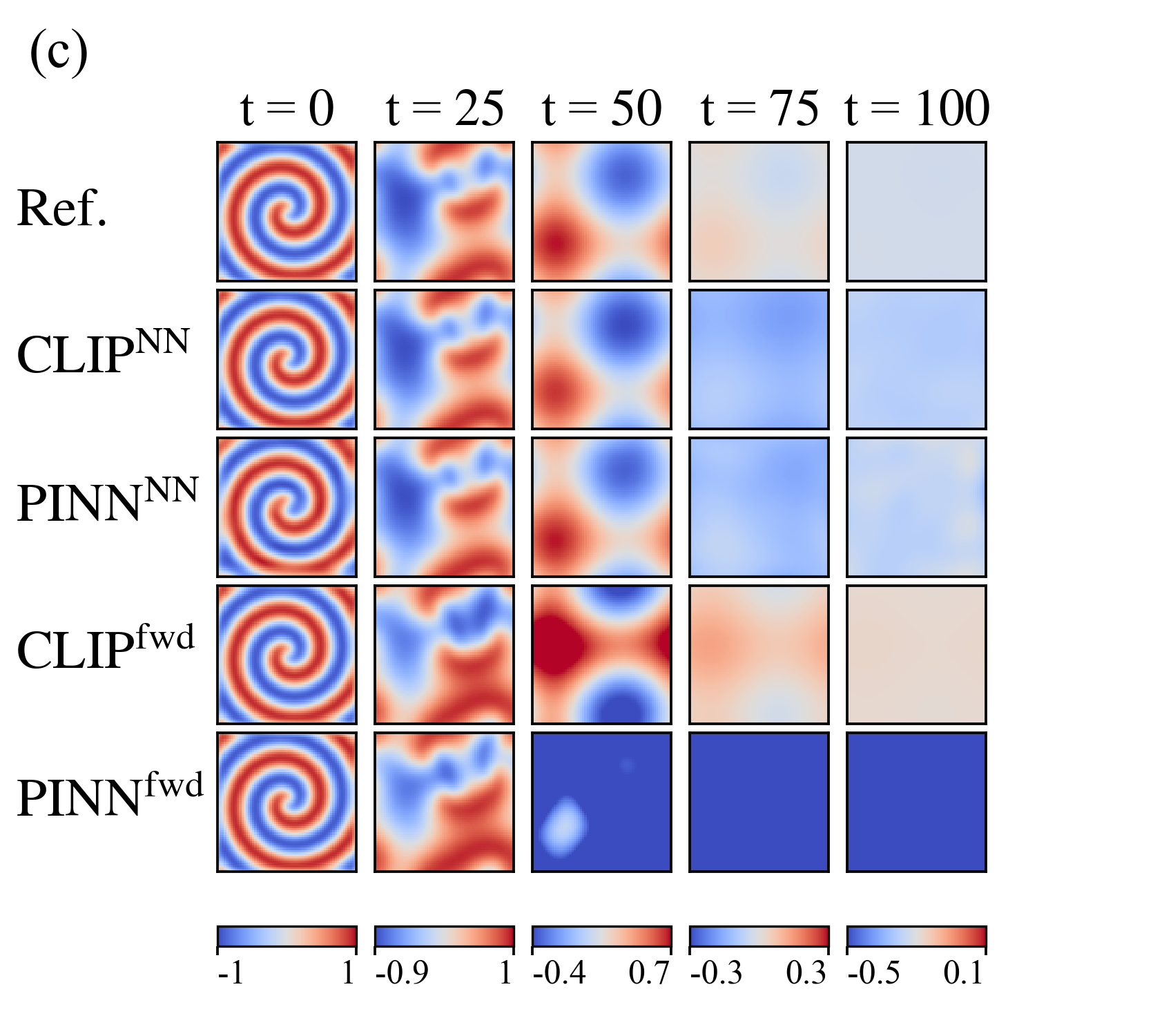}
    \end{minipage}%
    \hspace{0.014\textwidth}%
    \begin{minipage}[t]{0.4807\textwidth}
        \vspace{0pt}
        \includegraphics[width=\linewidth]
        {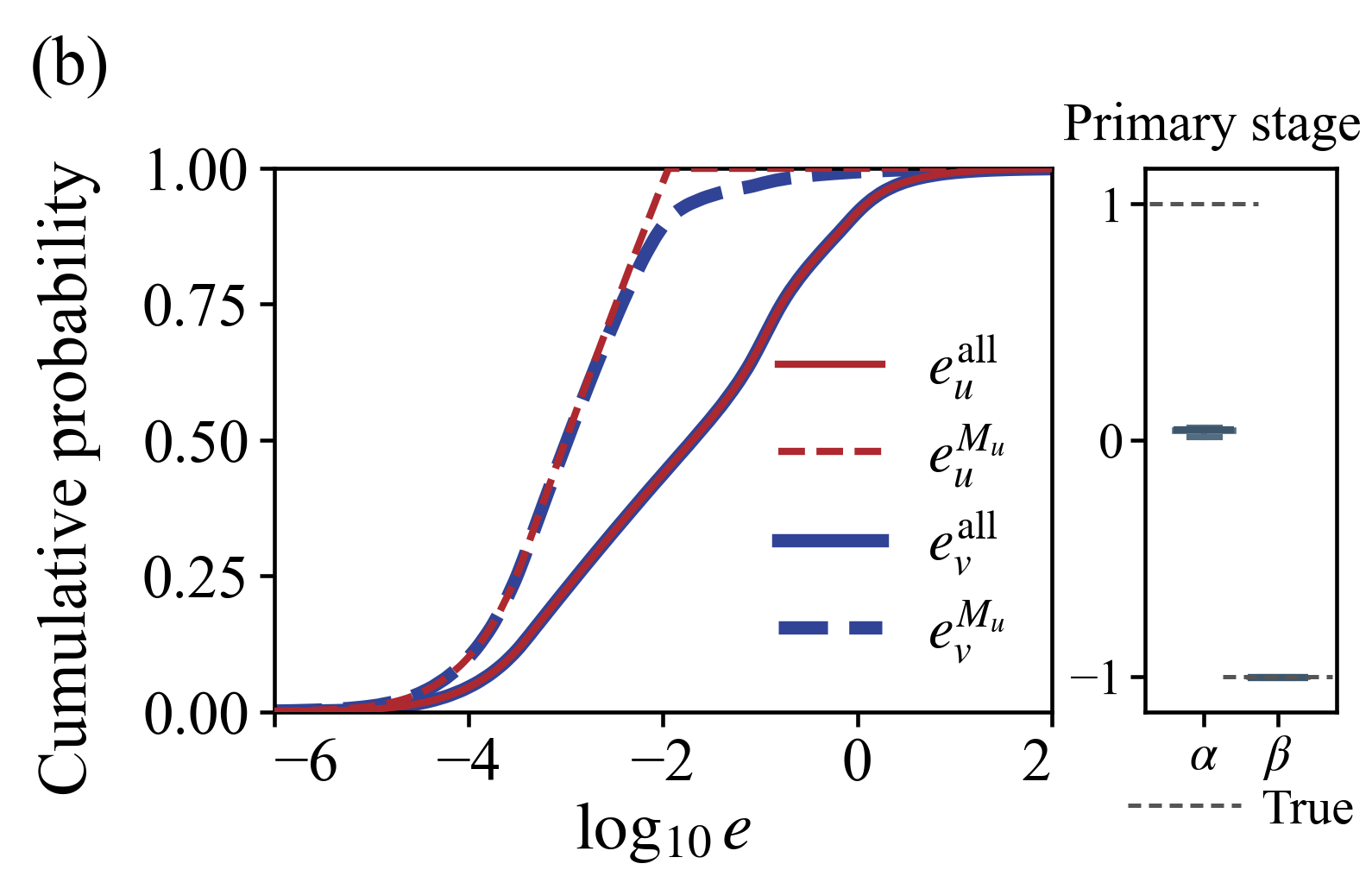}\par
        \nointerlineskip
        \includegraphics[width=\linewidth]
        {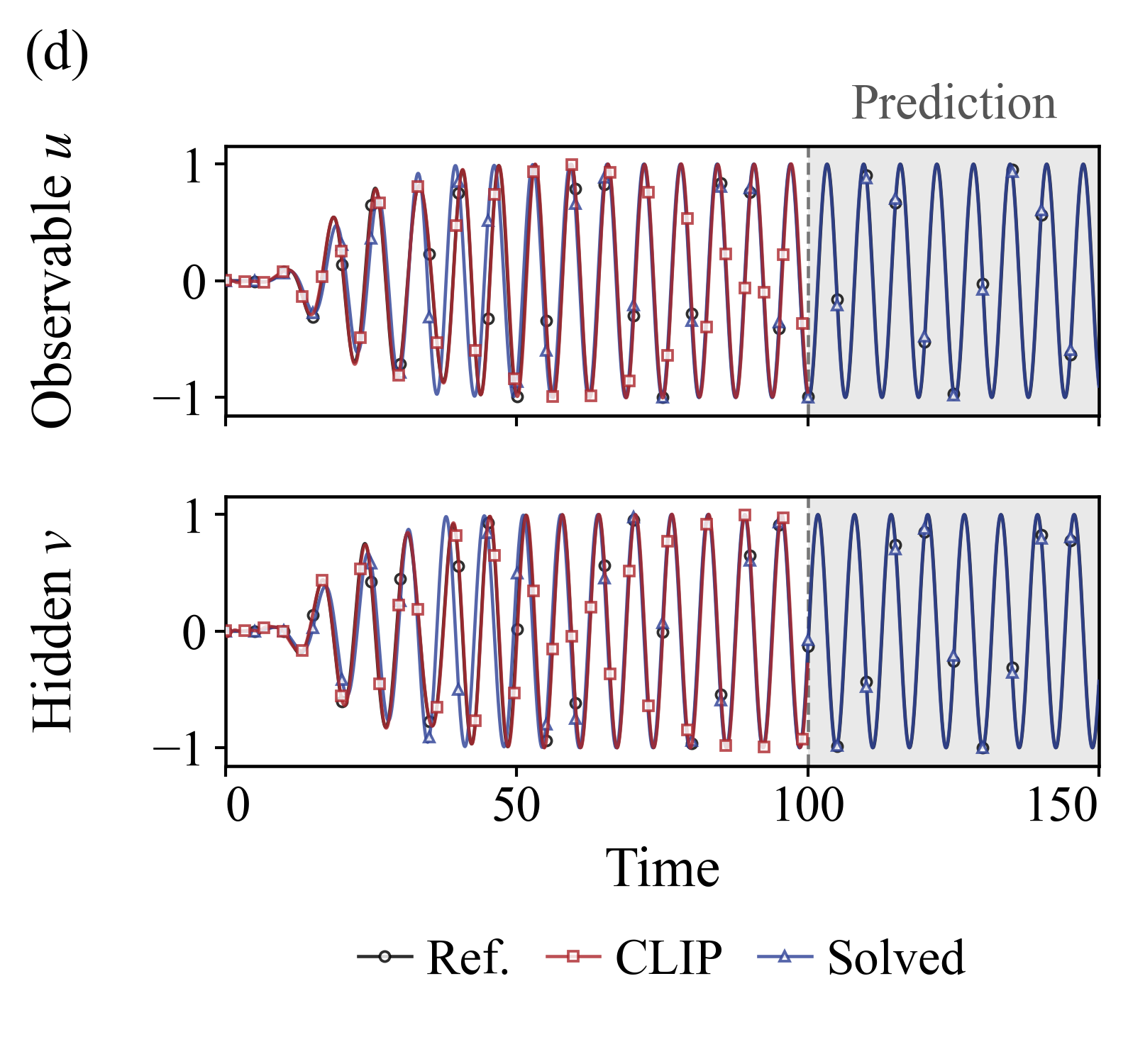}
    \end{minipage}
\caption{Results for the $\lambda$--$\omega$ reaction--diffusion system.
(a) Reaction-dominated mask $M_u$ and snapshots of the
reaction-approximation $e_u$ and $e_v$.
(b) Empirical cumulative distribution functions of $e_u$ and $e_v$
over the full space--time domain and within $M_u$, together with
estimates of $\alpha$ and $\beta$ obtained in the primary stage
over five runs.
(c) Spatial reconstructions and forward numerical solutions
of the unobserved field $v$.
The superscripts $\mathrm{NN}$ and $\mathrm{fwd}$ denote
direct neural network reconstructions and forward numerical
solutions obtained with the identified parameters, respectively.
(d) Pointwise trajectories of the observed state $u$
and the unobserved state $v$. The shaded regions indicate the prediction interval.}
\label{fig:lambda_reconstruction}
\end{figure}

\begin{table}[!htbp]
\centering
\caption{Mean relative absolute errors of the identified parameters
for CLIP and baseline methods across three reaction--diffusion systems
at different noise levels.}
\vspace{6pt}
\label{tab:noisy_results}
\setlength{\tabcolsep}{1pt}
\renewcommand{\arraystretch}{1.05}
\small
\begin{tabular*}{\textwidth}{@{\extracolsep{\fill}}lccccccccc@{}}
\toprule
& \multicolumn{3}{c}{\makebox[0pt][c]{$\lambda$-$\omega$ RD}} 
& \multicolumn{3}{c}{\makebox[0pt][c]{Gray-Scott}} 
& \multicolumn{3}{c@{}}{\makebox[0pt][c]{Lotka-Volterra}} \\
\cmidrule(lr){2-4}\cmidrule(lr){5-7}\cmidrule(lr){8-10}
Noise
& 0 & 5\% & 10\%
& 0 & 5\% & 10\%
& 0 & 5\% & 10\% \\
\midrule
\textbf{CLIP} & 6.56\% & 8.09\% & 16.50\% & 32.60\% & 37.71\% & 43.02\% & 13.6\% & 17.4\% & 16.8\% \\
PINN         & 7.20\% & 8.49\% & 14.32\% & 42.55\% & 88.63\% & 67.54\% & 30.3\% & 31.5\% & 33.4\% \\
PSO          & 153.1\% & 146.8\% & 351.2\% & 2312.4\% & \textemdash & \textemdash & 1215.8\% & 1215.5\% & 1217.1\% \\
ENKF         & 0.5\% & \textemdash & \textemdash & \textemdash & \textemdash & \textemdash & \textemdash & \textemdash & \textemdash \\
\bottomrule
\end{tabular*}
\par\vspace{5pt}

{\footnotesize\raggedright
\emph{Note.}\quad
Results for CLIP, PINN, and ENKF are averaged over five repeated runs
for each experimental condition.
PSO results are obtained from a single run for each condition.
An em dash indicates that the corresponding method did not converge.
\par}
\end{table}

Across the nine system and noise combinations in Table~\ref{tab:noisy_results}, CLIP gives a lower mean parameter MRAE than the standard PINN in eight. The exception is the λ–ω system at $10\%$ noise, where the MRAEs are $16.50\%$ for CLIP and $14.32\%$ for PINN.
Figs.~\ref{fig:lambda_reconstruction}(a--b) show the reaction-dominated regions selected by the observation mask $M_u$ and the parameter estimates obtained in the primary stage.
 The fraction of points satisfying $e_v<0.01$ increases from $43.8\%$ over the full space–time domain to $89.4\%$ within $M_u$. 
However, the primary-stage estimate of $\alpha$ is close to zero, whereas $\beta$ is estimated more accurately. 
The mask mainly selects later-time regions where the amplitude has nearly stabilized at $A\approx1$. 
For the amplitude $A=u^2+v^2$ and phase $\phi=atan2(v,u)$, the local reaction approximation at a fixed spatial location with $A>0$ gives 
\begin{equation}
\dot{A} = 2u\dot{u} + 2v\dot{v} = 2\alpha A(1 - A), \qquad \dot{\phi} = \frac{u\dot{v} - v\dot{u}}{A} = \beta A.
\end{equation}
In these regions, the amplitude-relaxation terms containing $\alpha$ are weak, whereas phase evolution remains sensitive to $\beta$. 
This is consistent with the observed difference in parameter estimates and indicates that reaction-dominated selection does not ensure equally informative constraints on all reaction parameters. 
Subsequent training over the full space–time domain includes the amplitude transients and provides additional constraints on $\alpha$.

The reconstruction of the unobserved field $v$ and the pointwise trajectories obtained by CLIP are compared with the reference solution in Fig.~\ref{fig:lambda_reconstruction}(c--d), showing agreement in the main spatiotemporal features.
The forward numerical solutions obtained with the identified parameters also capture the dominant oscillatory behavior, with comparable amplitudes and broadly consistent phase relationships.
In this relatively simple setting, a standard PINN already provides strong performance, with overall parameter errors comparable to those of CLIP, whereas PSO yields substantially larger parameter errors under partial observations.
EnKF has the lowest MRAE with clean data but does not converge in the tested noisy settings, as reported in Table~\ref{tab:noisy_results}.
These results provide a basic validation of joint parameter identification and hidden-state reconstruction under partial observations.

\begin{figure}[!htbp]
    \centering
    \includegraphics[width=0.505\linewidth]{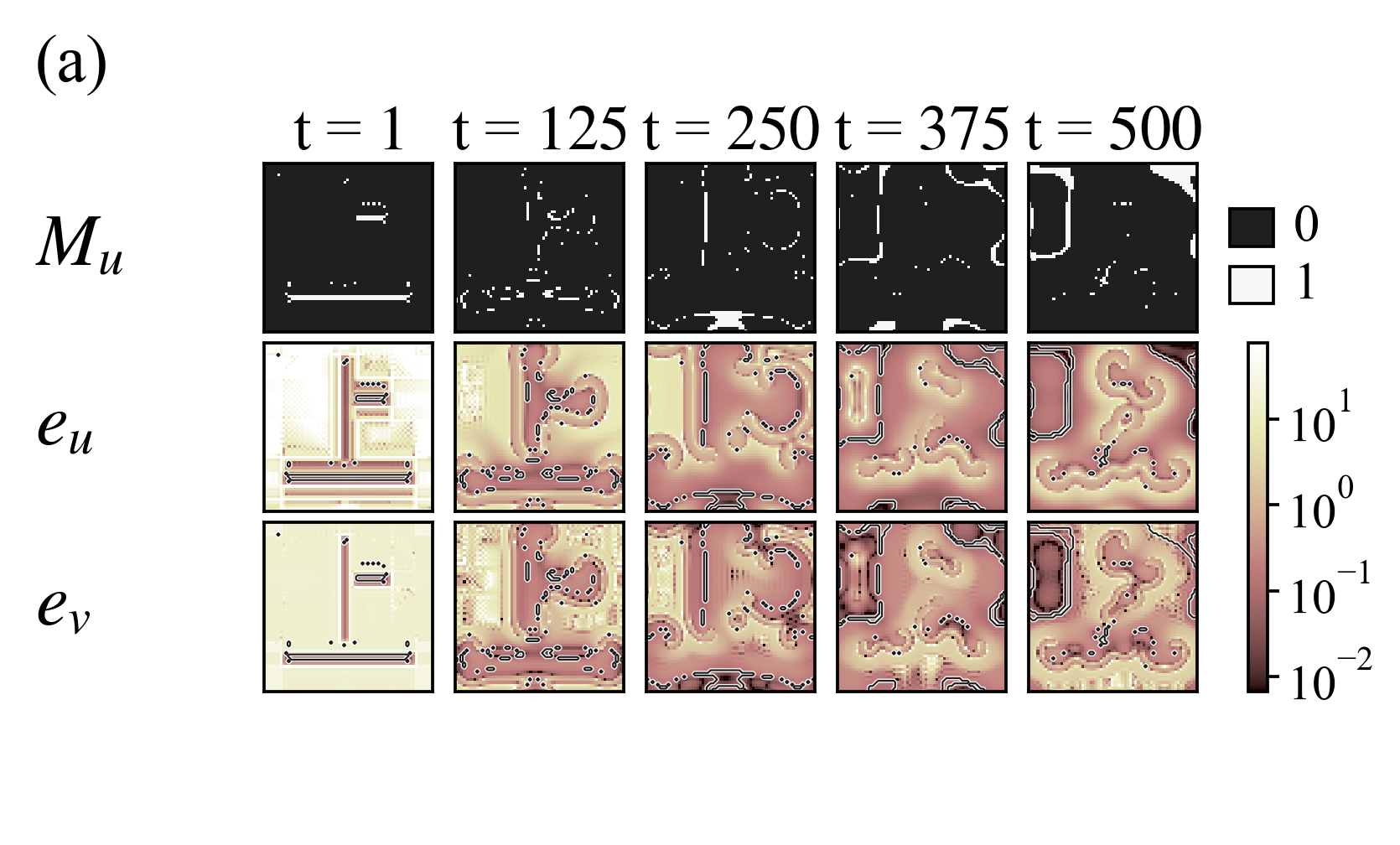}%
    \hspace{0.014\linewidth}%
    \includegraphics[width=0.4807\linewidth]{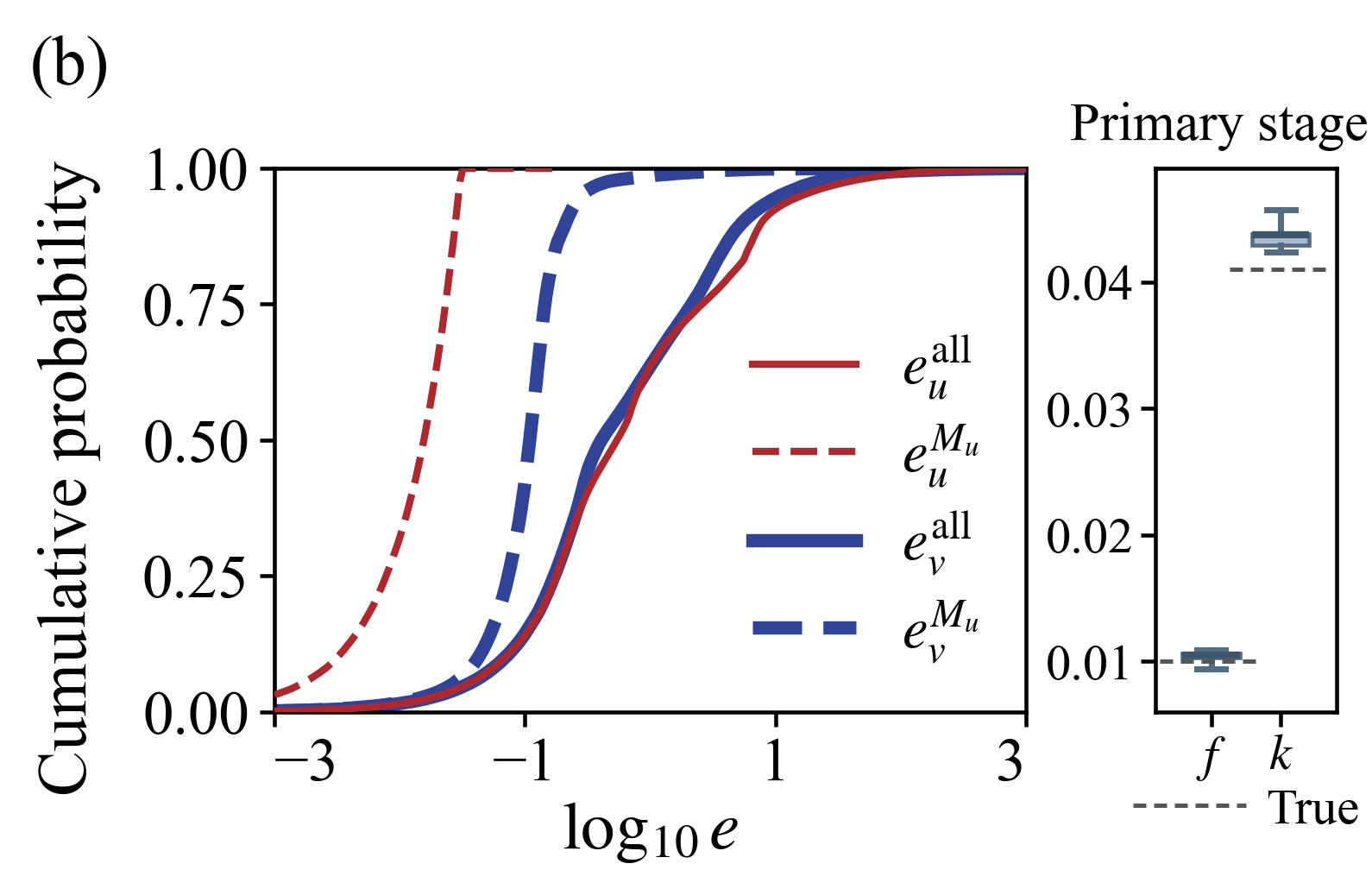}
    \caption{ Reaction-region selection and reaction-parameter estimation in the Gray--Scott system. 
(a) Reaction-dominated mask $M_u$ and snapshots of the
reaction-approximation $e_u$ and $e_v$.
(b) Empirical cumulative distribution functions of $e_u$ and $e_v$ over the full space--time domain and within $M_u$, together with estimates of $f$ and $k$ obtained in the primary stage over five runs.
}
    \label{fig:gray_reaction_selection}
\end{figure}

\begin{figure}[!htbp]
    \centering

    \resizebox*{!}{0.82\textheight}{%
        \begin{minipage}{\linewidth}
            \centering

            \begin{minipage}[t]{0.503\linewidth}
                \vspace{0pt}
                \includegraphics[width=\linewidth]
                {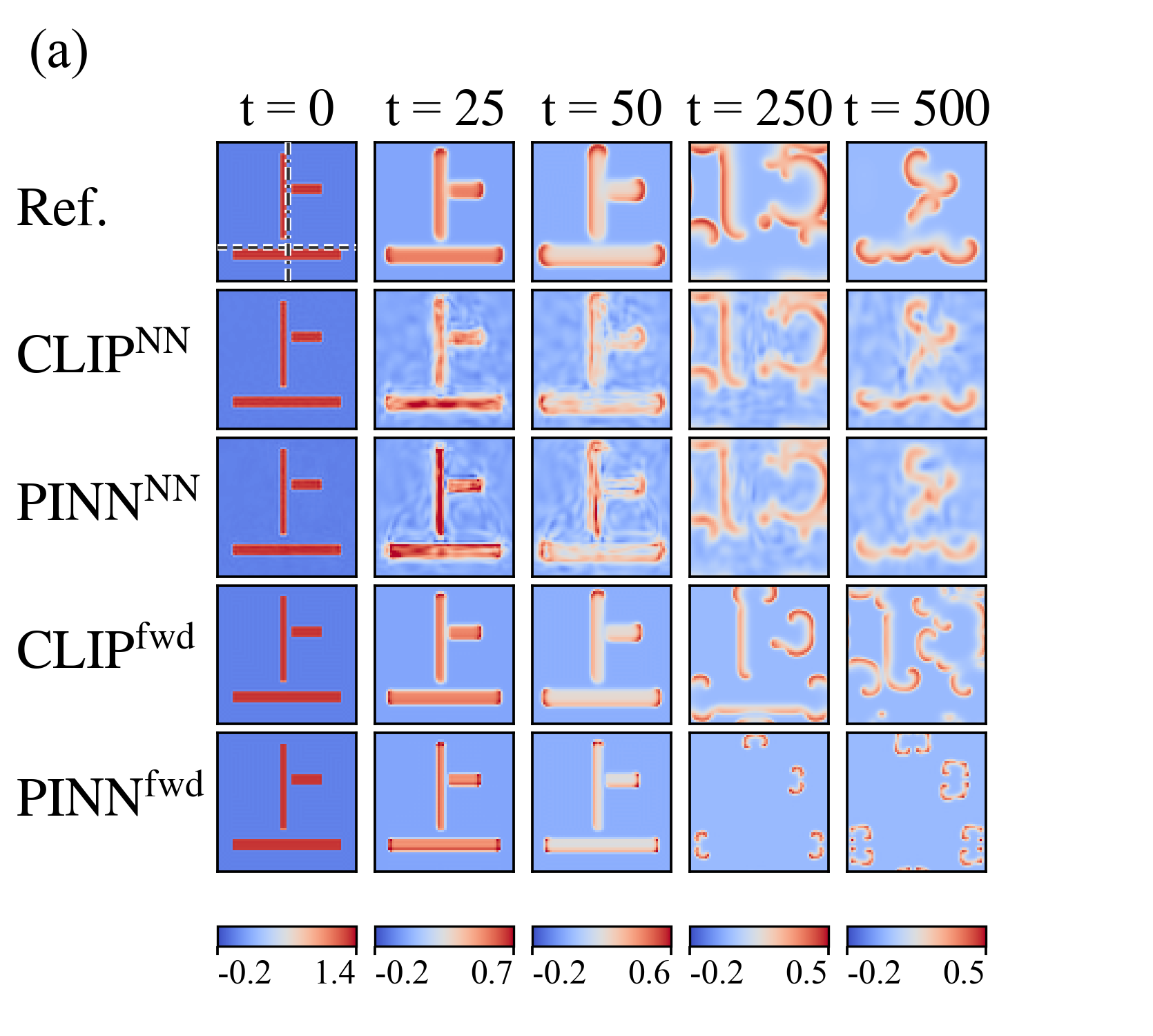}
            \end{minipage}%
            \hfill
            \begin{minipage}[t]{0.479\linewidth}
                \vspace{0pt}
                \includegraphics[width=\linewidth]
                {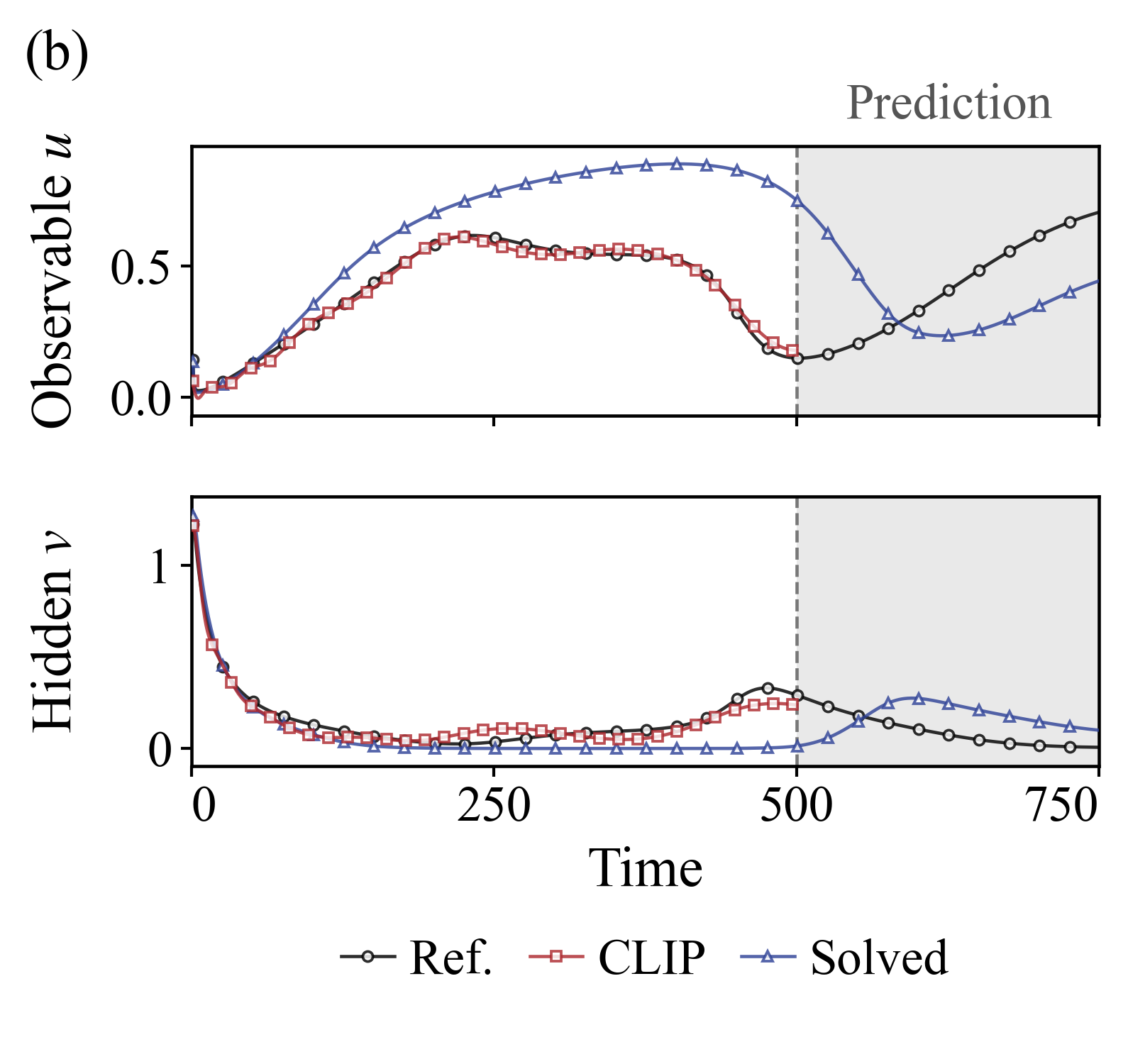}
            \end{minipage}

            \par\nointerlineskip
            \vspace{0mm}
            \includegraphics[width=\linewidth]
            {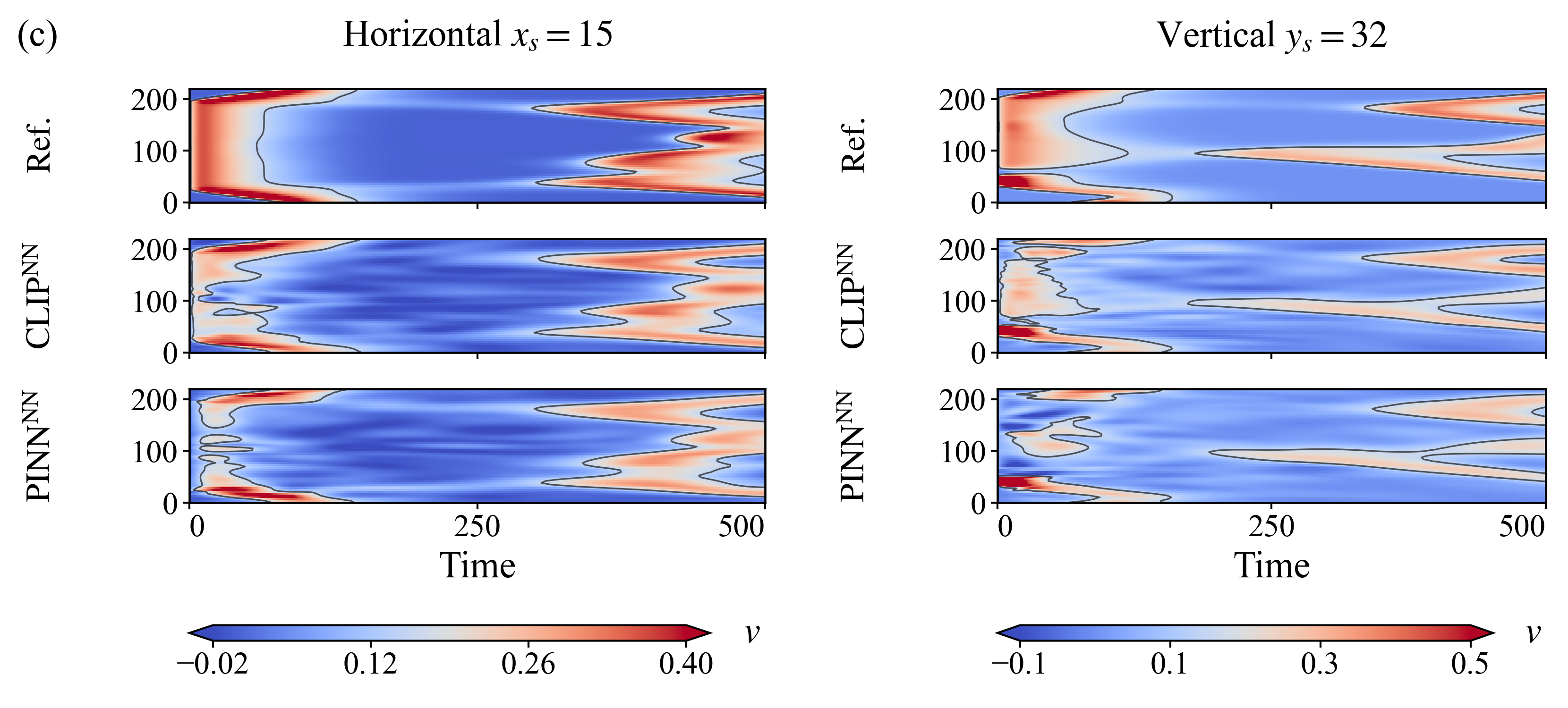}

            \par\nointerlineskip
            \vspace{0mm}
            \includegraphics[width=\linewidth]
            {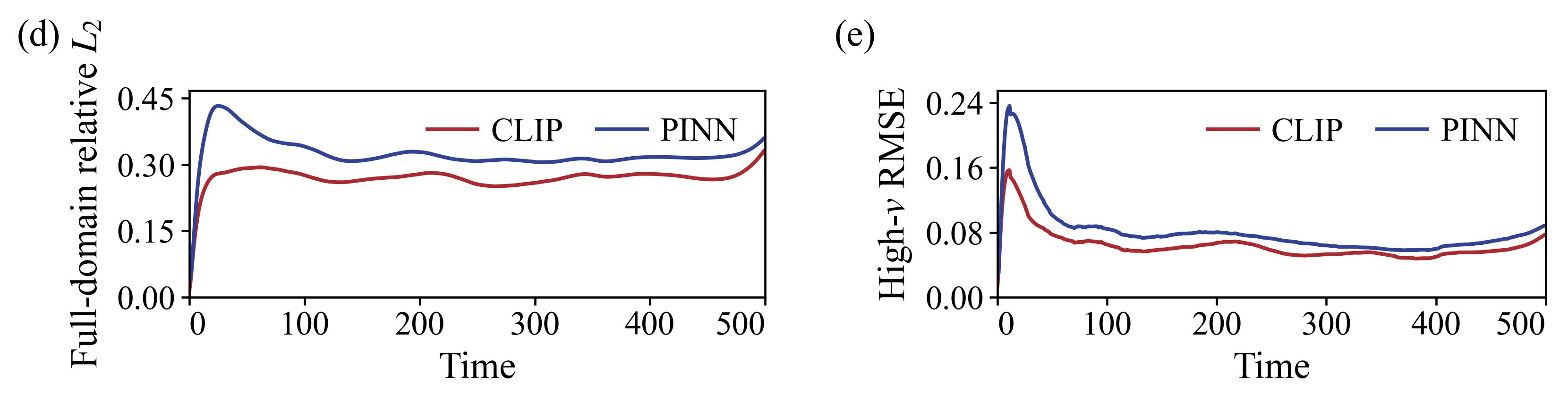}
        \end{minipage}%
    }

    \caption{Hidden-state reconstruction and dynamical comparisons
    for the Gray--Scott system.
(a) Spatial reconstructions and forward numerical solutions of
the unobserved field $v$.
(b) Pointwise trajectories of the observed state $u$ and the
unobserved state $v$.
(c) Space--time comparisons along the horizontal cut $x_s=15$
and vertical cut $y_s=32$, marked in the initial reference
snapshot in (a). Black contours indicate $v=0.15$.
(d) Time-resolved full-domain relative $L_2$ errors of the
CLIP and PINN reconstructions of $v$.
(e) Root-mean-square errors of the reconstructed $v$ in
high-concentration regions where $v_{\mathrm{ref}}\geq0.15$.}
    \label{fig:gray_reconstruction}
\end{figure}

\subsection{Case 2: The Gray-Scott equations}
We next examine the Gray-Scott reaction-diffusion system, a prototypical model for chemical pattern formation. Depending on the parameter regime, it generates diverse spatiotemporal structures, including spots, stripes, and self-replicating localized structures \cite{pearson1993gray}. Its nonlinear reaction kinetics and parameter-sensitive dynamics make it a challenging setting for joint parameter estimation and hidden-field reconstruction. We consider the two-dimensional system with periodic boundary conditions,
\begin{equation}
u_t = D_1 \Delta u - u v^2 + f(1-u),
\end{equation}
\begin{equation}
v_t = D_2 \Delta v + u v^2 - (f+k)v,
\end{equation}
where $u$ and $v$ denote the state variables.
The unknown parameters include the diffusion coefficients $D_1$ and $D_2$, and the feed and removal rates $f$ and $k$.
In this experiment, only the field $u$ is observed, and we infer $(D_1,D_2,f,k)$ while reconstructing the unobserved field $v$.
The nonlinear coupling through $uv^2$ links parameter estimation to hidden-field reconstruction, while localized spatial structures require accurate resolution of spatial variations in $v$.

Figure~\ref{fig:gray_reaction_selection} shows the candidate reaction-dominated regions selected for reaction-parameter initialization.
Both high-gradient regions near wavefronts and extensive near-zero regions with little information about the reaction kinetics are excluded.
The remaining informative regions therefore occupy only a limited spatial extent.
Despite this restriction, the fraction of points satisfying $e_v<0.1$ increases from $14.35\%$ before selection to $42.11\%$ among the selected points (Fig.~\ref{fig:gray_reaction_selection}b).
Reaction-parameter estimation within these regions yields first-stage estimates of $f$ and $k$ close to their reference values.
Thus the observed-field mask enriches the selected sample in points with weak diffusion in $v$.
The resulting estimates serve as initial values for subsequent full PDE training.

The neural-network reconstruction in Fig.~\ref{fig:gray_reconstruction}a broadly captures the spatial morphology of the hidden field $v$.
Figure~\ref{fig:gray_reconstruction}c further shows that CLIP captures the early evolution in two representative space--time slices with pronounced structural variation.
During reconstruction, CLIP generally gives a lower full-domain relative $L_2$ error in $v$ than the standard PINN. The reduction also extends to the RMSE in the high-$v$ regions. 
The first-stage estimates of $f$ and $k$ in Fig.~\ref{fig:gray_reaction_selection}b have small errors.
At the end of training, Table~\ref{tab:noisy_results} reports the mean error over all four coefficients.
The hidden field $v$ enters the reaction kinetics through $uv^2$, making the reaction term sensitive to errors in its reconstructed amplitude.
The underestimated peak amplitudes weaken the reaction term, and this bias may be partly offset by changes in the diffusion terms or reaction parameters.
Nonuniform errors in near-zero regions can also distort $\Delta v$, and a smaller $D_2$ may reduce the contribution of these derivative errors to the diffusion term.
As shown in Fig.~\ref{fig:gray_reconstruction}b, 
the reconstructed $v$ follows the rise towards the local peak shortly before $t = 500$, though it underestimates the peak height. The forward solution with the CLIP coefficients remains close to zero at this time and peaks later, during the prediction interval.
These results suggest that pretraining the reaction parameters alleviates part of the parameter identification difficulty, while accurate joint recovery of diffusion coefficients and hidden spatial structures remains challenging under partial observations.

\begin{figure}[!htbp]
    \centering
    \includegraphics[width=0.505\linewidth]{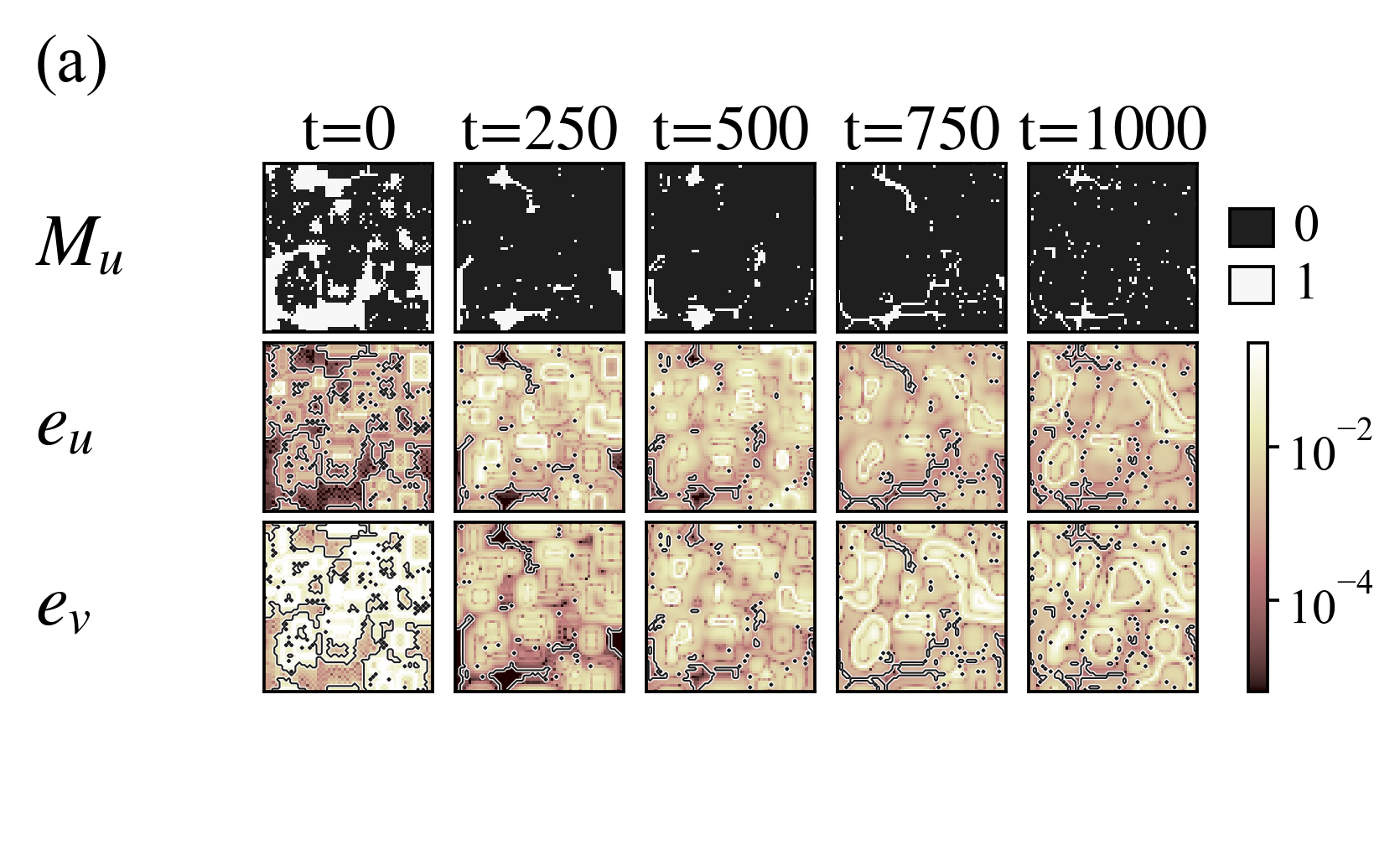}%
    \hspace{0.014\linewidth}%
    \includegraphics[width=0.4807\linewidth]{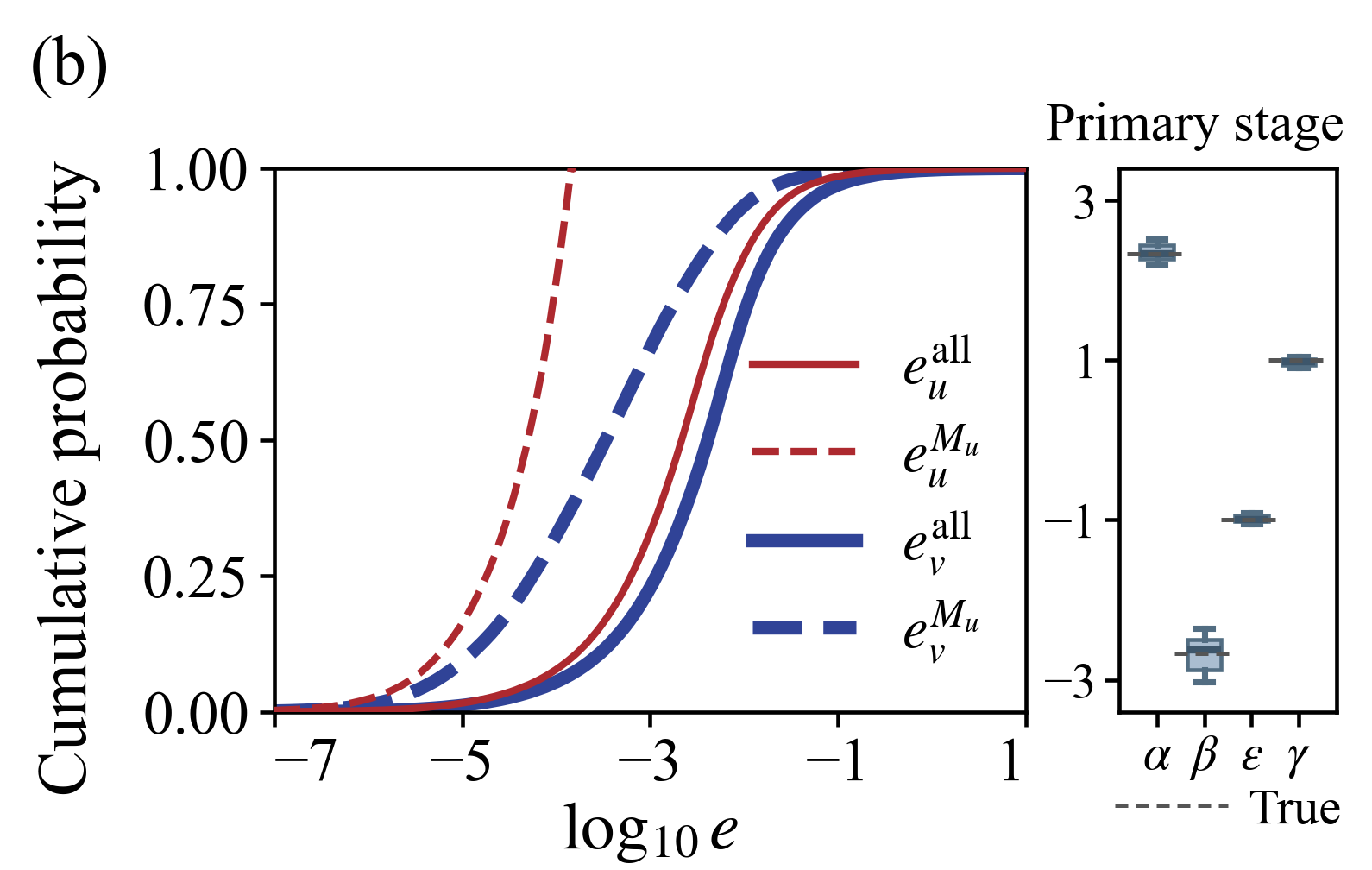}
    \caption{Reaction-region selection and reaction-parameter estimation
    in the Lotka-Volterra system. 
(a) Reaction-dominated mask $M_u$ and snapshots of the
reaction-approximation $e_u$ and $e_v$.
(b) Empirical cumulative distribution functions of $e_u$ and $e_v$ over the full space--time domain and within $M_u$, together with estimates of $\alpha, \beta, \varepsilon$ and $\gamma$ obtained in the primary stage over five runs.
    }
    \label{fig:lotka_reaction_selection}
\end{figure}

\subsection{Case 3: The Lotka-Volterra equations}
The Lotka-Volterra reaction-diffusion system is a classical model of spatial predator-prey dynamics that can generate traveling waves and heterogeneous population distributions. 
With six unknown coefficients and nonlinear coupling to an unobserved state, this system poses a challenging problem for joint parameter estimation and hidden-state reconstruction.
The two-dimensional governing equations are
\begin{equation}
u_t = D_1 \Delta u + \alpha u + \beta u v,
\end{equation}
\begin{equation}
v_t = D_2 \Delta v + \gamma u v + \epsilon v,
\end{equation}
with periodic boundary conditions.
Here, $u$ and $v$ denote the population densities of the two species.
The unknown coefficients include the diffusion rates $D_1$ and $D_2$, the intrinsic growth or decay rates $\alpha$ and $\epsilon$, and the interaction coefficients $\beta$ and $\gamma$.
Given observations of $u$ and the complete initial field $v_0$, the model estimates all six coefficients while reconstructing the subsequent evolution of the unobserved state $v$.

\begin{figure}[!htbp]
    \centering
    \makebox[\linewidth][l]{%
        \includegraphics[width=0.48\linewidth]{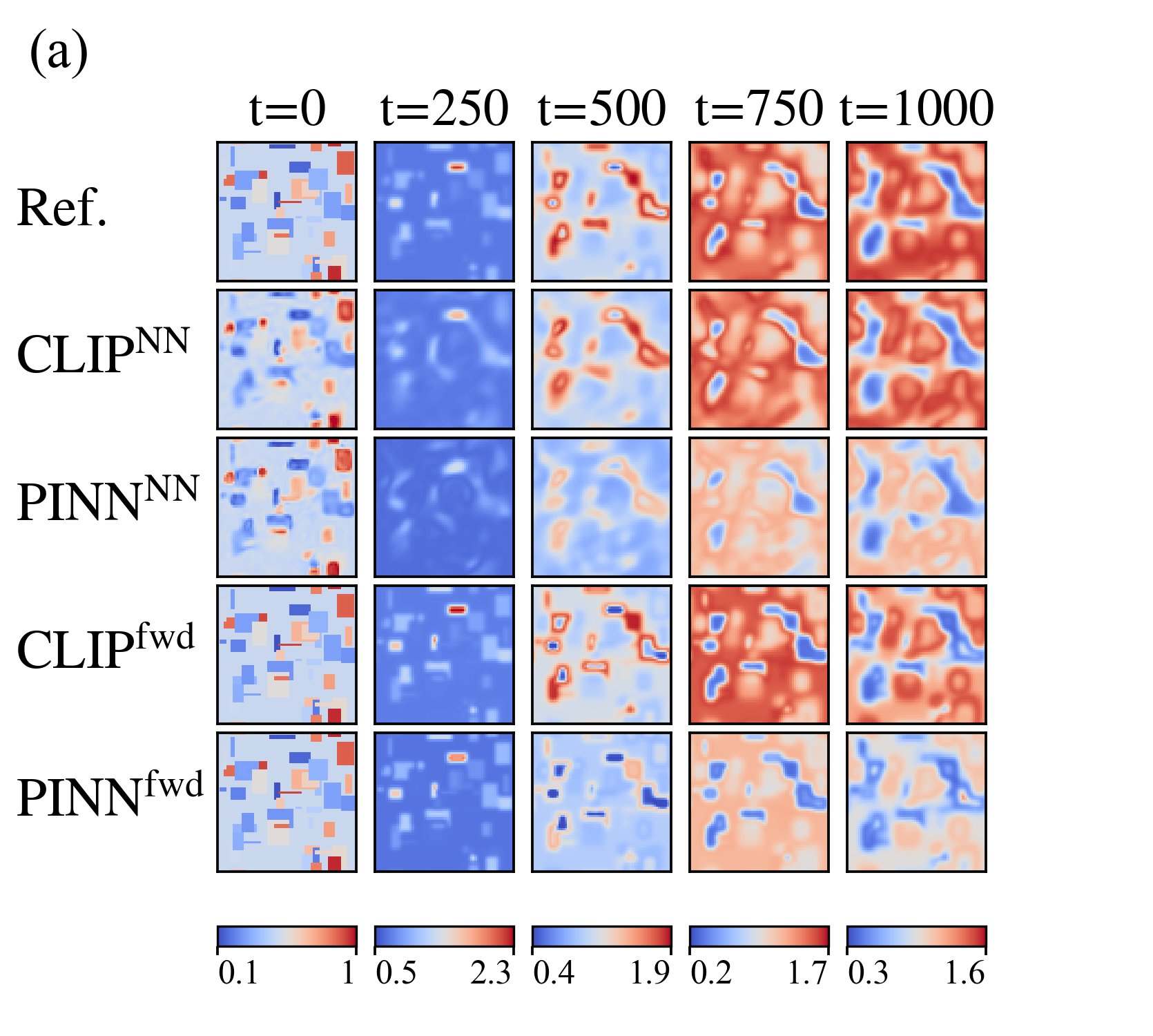}%
        \hspace{0.02\linewidth}%
        \includegraphics[width=0.457\linewidth]{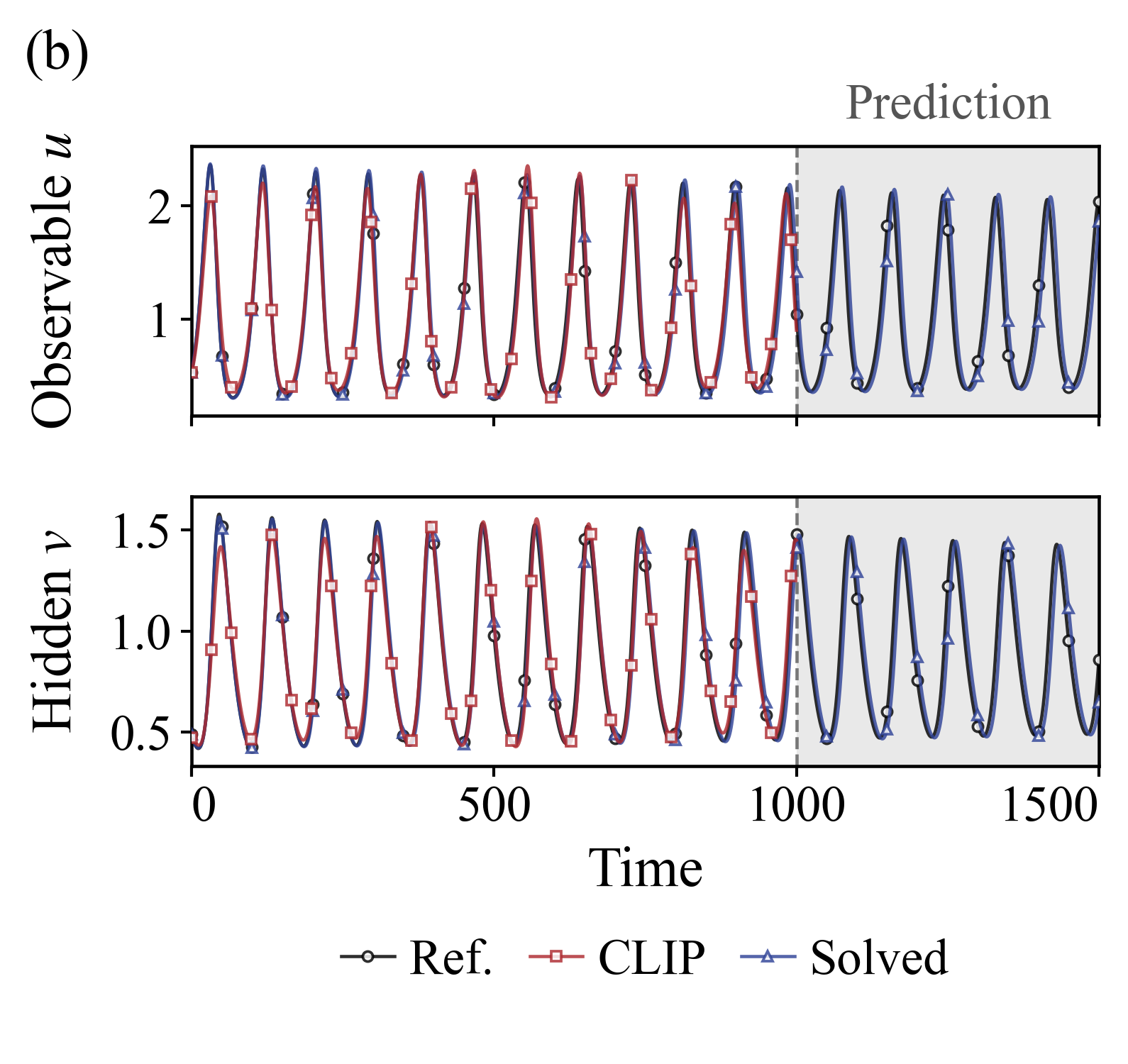}%
    }\par
    \vspace{1.5mm}
    \makebox[\linewidth][l]{%
        \includegraphics[width=0.48\linewidth]{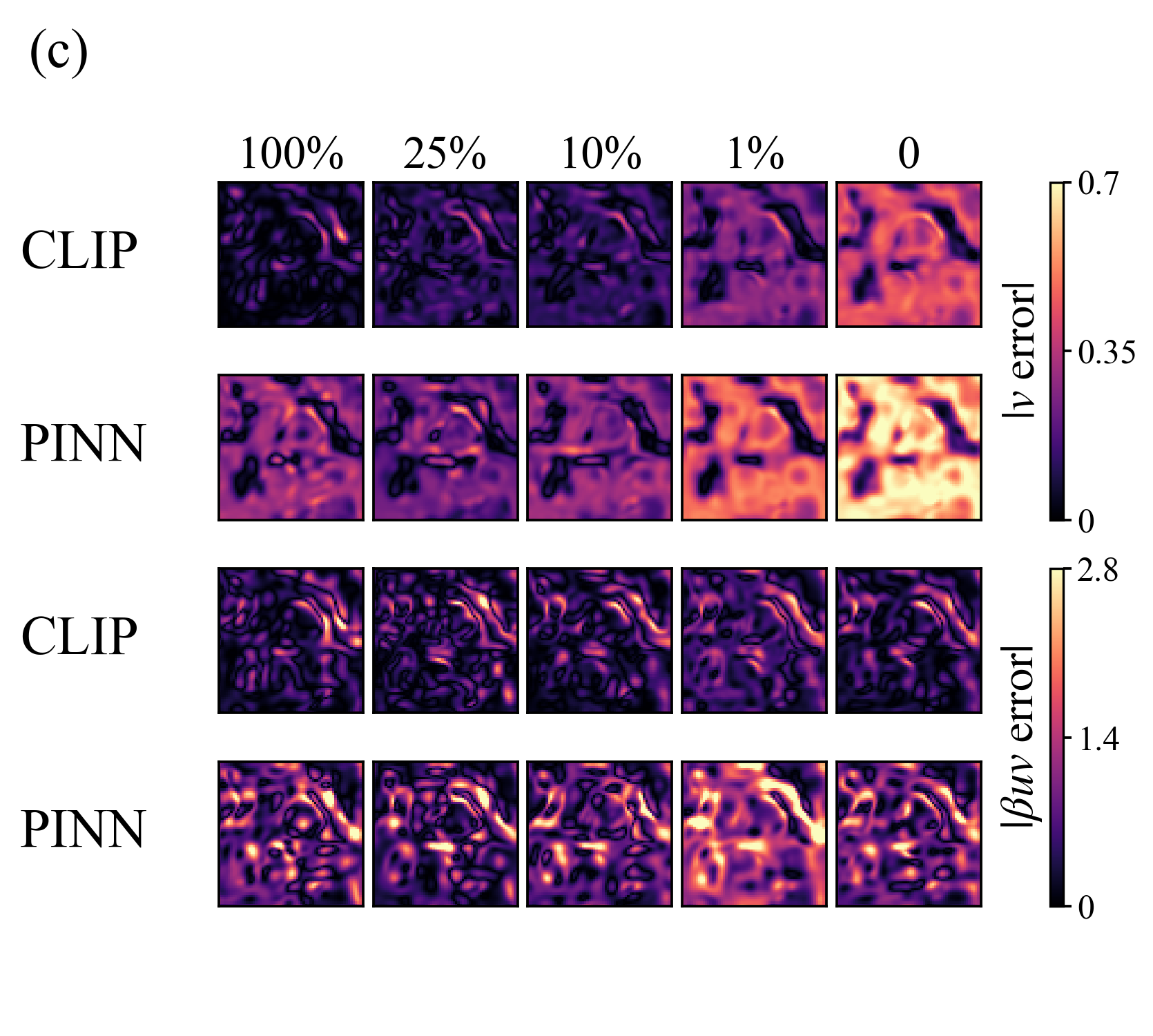}%
        \hspace{0.02\linewidth}%
        \includegraphics[width=0.48\linewidth]{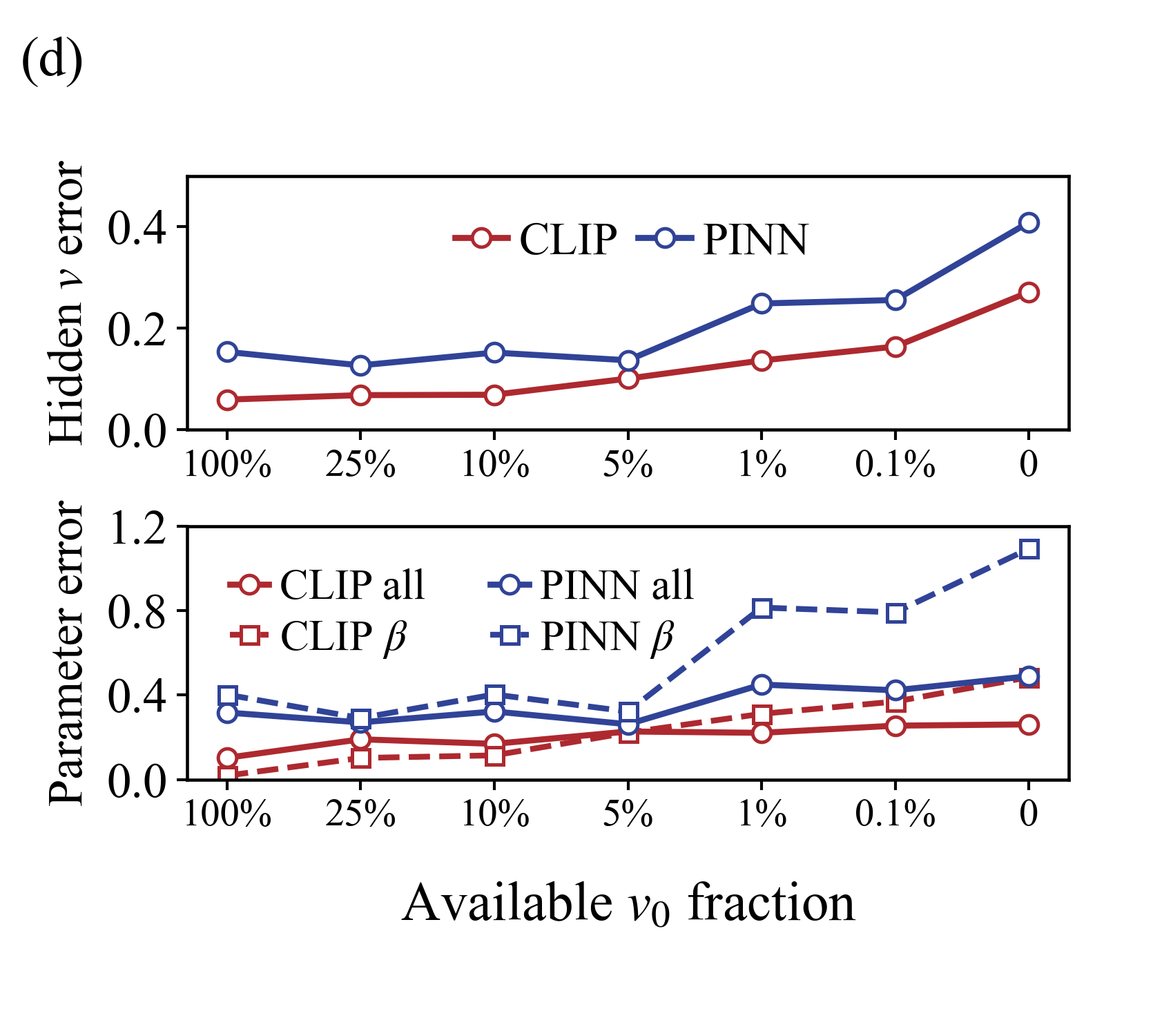}%
    }
    \caption{Spatial reconstruction, pointwise dynamics, and the effects of
    hidden initial-condition information in the Lotka--Volterra system.
(a) Spatial reconstructions and forward numerical solutions
of the unobserved field $v$.
(b) Pointwise trajectories of the observed state $u$
and the unobserved state $v$. The shaded regions indicate the prediction interval.
(c) Spatial absolute-error maps for CLIP and PINN under different observation fractions of the hidden initial field, showing errors in the reconstructed field $v$ (top) and the interaction term $\beta uv$ (bottom).
(d)  Hidden-state reconstruction and parameter estimation errors under different fractions of available hidden initial-condition observations.
The top panel shows the relative $L_1$ error in $v$ over the full space--time domain.
The bottom panel shows the MRAE
across all six parameters and the relative absolute error in $\beta$.
    }
    \label{fig:lotka_reconstruction}
\end{figure}

Figure~\ref{fig:lotka_reaction_selection} shows the region selection and reaction-parameter estimates obtained in the primary stage.
The mask $M_u$ is constructed from observations of $u$ to select reaction-dominated regions across space and time.
The proportion of samples satisfying $e_v < 0.01$ increases from 73.24\% over all samples to 93.15\% within the selected regions.
This enrichment indicates that selection based on the observed variable also favors regions where the hidden-state dynamics are reaction-dominated.
After pretraining on these regions, the reaction-parameter estimates are generally close to their reference values, although some deviations remain, as shown in Fig.~\ref{fig:lotka_reaction_selection}(b).
These results suggest that the primary stage provides a useful initial estimate of the reaction dynamics before diffusion is fully incorporated into the joint optimization of parameters and the hidden state.

Across all tested noise levels, CLIP achieves consistently lower identification errors, whereas PSO and EnKF yield larger errors in our experiments as shown in Table~\ref{tab:noisy_results}.
The reconstruction of the unobserved state $v$ also agrees more closely with the reference solution.
Figure~\ref{fig:lotka_reconstruction}(a) shows that the neural network output of CLIP captures the spatial distribution and temporal evolution of $v$, while the baseline PINN exhibits more pronounced amplitude errors.
Forward simulations using the estimated parameters show a similar advantage, indicating that the improvement also extends to the dynamics generated by the recovered coefficients.
At the sampled spatial location, the reconstructed time series of both $u$ and $v$ closely follow the reference trajectories, and the forward solution continues to track their oscillation amplitudes and phases over the subsequent prediction interval (Fig.~\ref{fig:lotka_reconstruction}(b)).
These results support the joint recovery of the hidden-state dynamics and the governing parameters.

When initial observations of the hidden state are unavailable and no other scale constraints are imposed, the Lotka--Volterra system admits a scaling ambiguity.
Multiplying $v$ by an admissible constant $c>0$ and dividing $\beta$ by $c$ leaves the evolution of $u$ and the interaction term $\beta uv$ unchanged.
Thus, observations of $u$ alone cannot uniquely determine the scales of $\beta$ and $v$.
This symmetry is explicit if we set $w = \beta v$. The equations become $u_t = D_1\Delta u + \alpha u + uw$ and $w_t = D_2\Delta w + (\gamma u + \varepsilon)w$. The evolution equations contain no separate factor of \(\beta\), and each admissible rescaling gives the same interaction term \(uw\). 
To examine the dependence of recovery on initial information, the fraction of observed values in $v_0$ is progressively reduced to zero, and both $v$ and the interaction term $\beta uv$ are evaluated.
Figures~\ref{fig:lotka_reconstruction}(c) and (d) show that recovery errors generally increase as initial observations become sparser, but CLIP maintains lower hidden-state and overall parameter errors across the tested fractions.
The baseline PINN also exhibits a larger increase in the estimation error of $\beta$ as initial information is reduced.
The reduced spatial errors in $\beta uv$ further indicate that the advantage of CLIP extends to the nonlinear interaction term in the observed equation (Fig.~\ref{fig:lotka_reconstruction}(c)).
These results support the effectiveness of staged training in joint parameter estimation and hidden-state reconstruction and are consistent with reduced error compensation during optimization.

\subsection{Case 4: Identification of high-dimensional biological systems}\label{sec-minde}
We consider MinD and MinE self-organization in \textit{Escherichia coli}, where pole-to-pole protein oscillations contribute to cell division site positioning.
Based on biochemical mechanisms and experimental observations, Loose et al.~\cite{loose2008minde} proposed a reaction--diffusion model coupling cytosolic and membrane-bound protein dynamics.
Compared with the previous canonical benchmarks, this model provides a more complex test of parameter identification with biophysically interpretable kinetic and diffusion coefficients.
The governing equations are
\begin{equation}
\begin{aligned}
& \partial_t c_D = \omega_{de}c_{de} - c_D(\omega_D + \omega_{dD}c_d)+D_D\Delta  c_D,\\
& \partial_t c_E = \omega_{de}c_{de} -c_Ec_d(\omega_E + \omega_{eE}c_{de}^2) + D_E\Delta c_E,\\
& \partial_t c_d = c_D(\omega_D + \omega_{dD}c_d) - c_Ec_d(\omega_E +\omega_{eE}c_{de}^2) + D_d\Delta c_d,\\
& \partial_t c_{de} = c_Ec_d(\omega_E + \omega_{eE}c_{de}^2) -\omega_{de}c_{de} + D_{de}\Delta  c_{de},
\end{aligned}
\label{eq:minde}
\end{equation}
where $c_D$ and $c_E$ denote the cytosolic concentrations of MinD and MinE, and $c_d$ and $c_{de}$ denote the concentrations of membrane-bound MinD and the MinD--MinE complex, respectively.
We infer five kinetic rates $(\omega_D,\omega_{dD},\omega_E,\omega_{eE},\omega_{de})$ and four diffusion coefficients $(D_D,D_E,D_d,D_{de})$ from membrane observations, while jointly reconstructing the cytosolic states from their prescribed initial fields.

\begin{figure}[!htbp]
    \centering
    \includegraphics[width=0.562\linewidth]{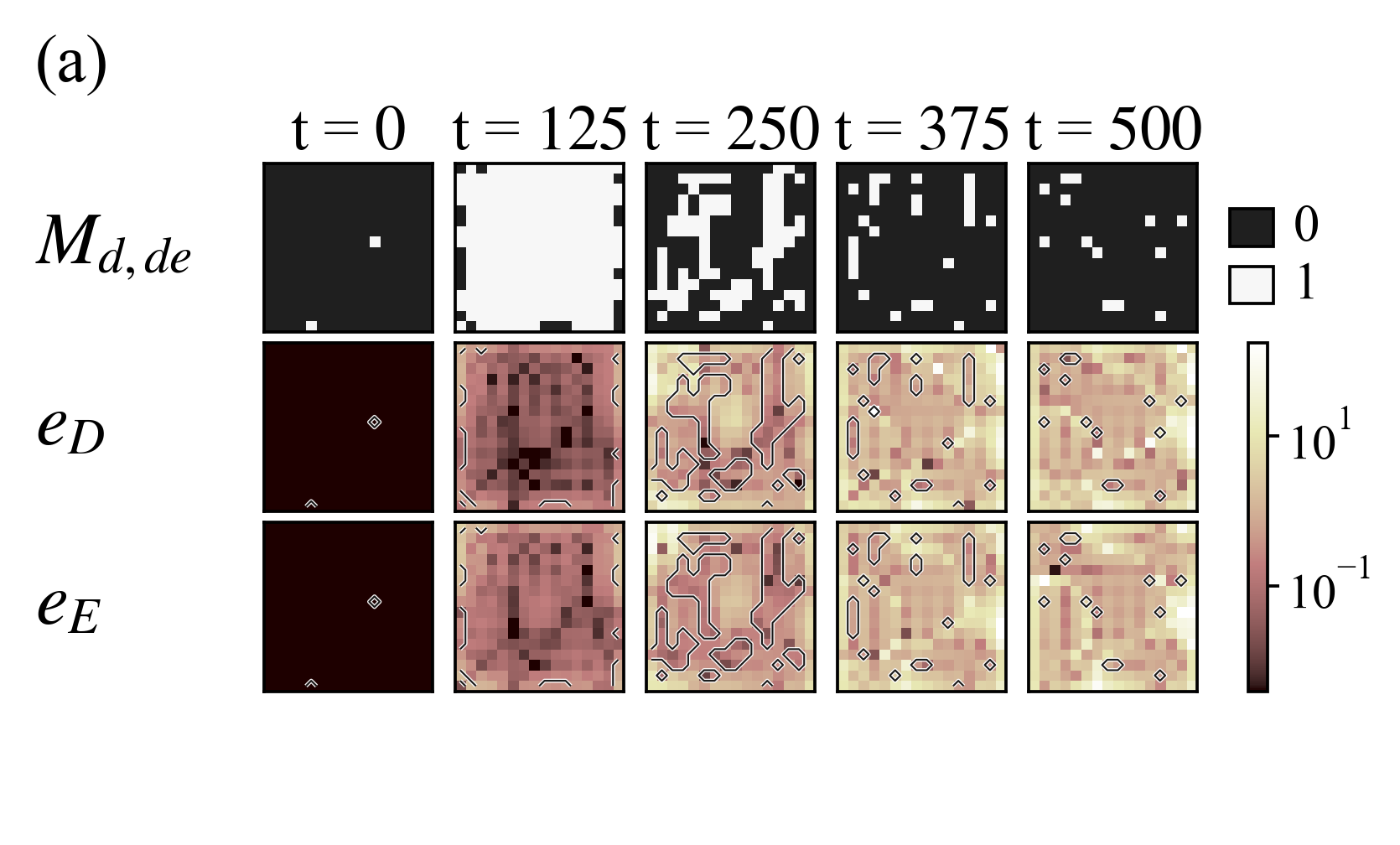}%
    \hspace{0.014\linewidth}%
    \includegraphics[width=0.424\linewidth]{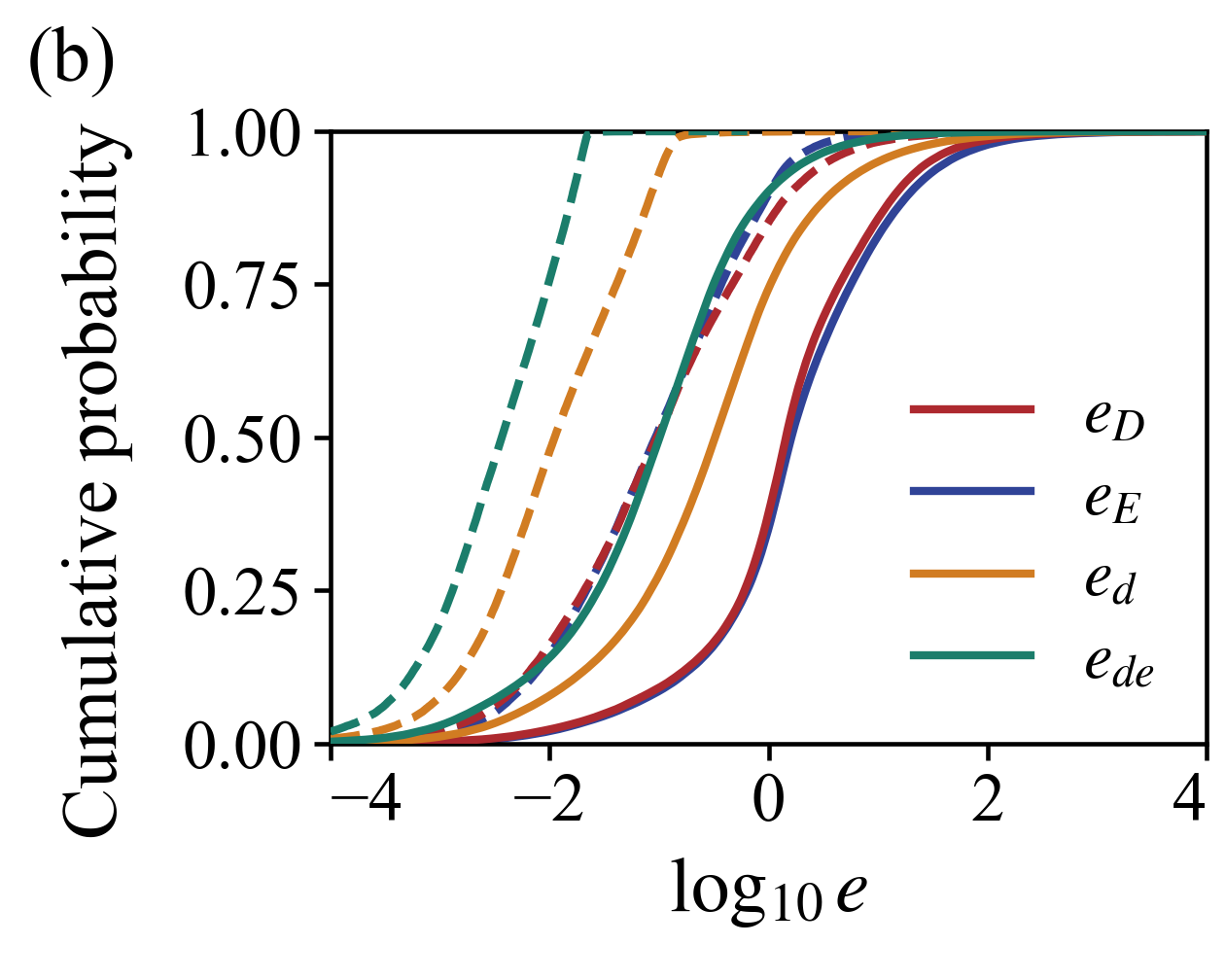}
    \caption{Reaction-region selection and diffusion-to-reaction ratio distributions in the MinDE system.
(a) The observation-based mask $M_{d,de}$ and the diffusion-to-reaction ratios $e_D$ and $e_E$ for the hidden cytosolic states at selected times. (b) Empirical cumulative distribution functions of $e_D, e_E, e_d$, and $e_de$. Solid curves represent all space–time points, whereas dashed curves represent points within $M_{d,de}$.
}
    \label{fig:min_reaction_selection}
\end{figure}

We train the network on membrane observations from a $16\times16$ local subregion at the same spatial resolution as the full simulation.
No boundary constraints are imposed at the edges of this interior subregion.
All kinetic rates and diffusion coefficients are optimized in log space to accommodate their different orders of magnitude.
Concentrations are normalized for training, while PDE residuals are evaluated on the original physical scale.
The identified parameters are then used in numerical simulations from new initial fields to assess their ability to reproduce the spatial patterns of both cytosolic species.

\begin{figure}[!htbp]
    \centering
    \includegraphics[width=\linewidth]{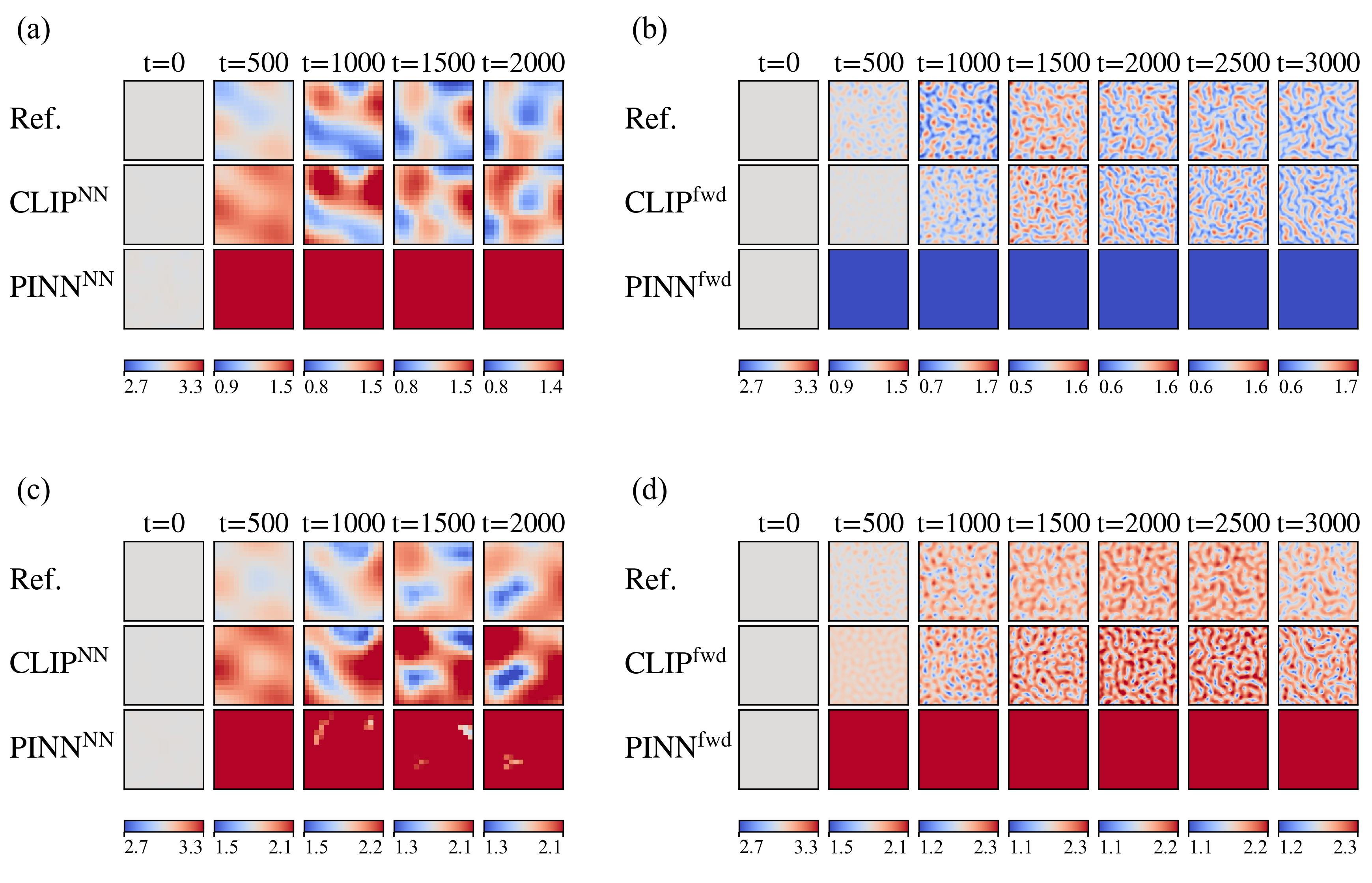}
    \caption{Hidden-state reconstruction and forward simulation in the
    MinDE system. 
(a,c) Spatial reconstructions of the unobserved fields $c_D$ and $c_E$.
(b,d) Forward numerical solutions for $c_D$ and $c_E$ under more complex
initial conditions using the identified coefficients.
All concentrations are divided by $10^6$.}
    \label{fig:min_reconstruction_combined}
\end{figure}

We first examine whether the mask constructed from membrane observations also enriches regions with small $e$ for the hidden cytosolic species, as shown in Fig.~\ref{fig:min_reaction_selection}.
The fractions of points satisfying $e<0.1$ increase from $9.13\%$ to $50.67\%$ for $c_D$ and from $8.58\%$ to $51.45\%$ for $c_E$, indicating enrichment for both hidden species.
Table~\ref{tab:min_results} reports the nine identified parameters.
On clean data, CLIP achieves a relative absolute error of $30.99\%$, whereas the baseline PINN
reaches $67.28\%$ and shows unstable optimization. 
With $5\%$ and $10\%$ additive Gaussian noise, CLIP main-
tains an RAE of $31.76\%$ and $25.71\%$, indicating robustness in this high dimensional inverse problem. 
CLIP recovers the main spatial structures of the cytosolic fields as shown in Fig.~\ref{fig:min_reconstruction_combined}(a,c) and yields parameter estimates that are overall closer to the reference values, but some deviations remain.
These may reflect the indirect constraints provided by local membrane observations and the dependence of parameter estimation on the reconstruction of hidden states.
By comparison, the baseline PINN fails to recover the corresponding spatial structures and severely underestimates both cytosolic diffusion coefficients.

\begin{table}[!htbp]
\caption{Parameter identification results for the \textit{E.~coli} Min protein self-organization model. CLIP estimates are compared against the reference parameter set.}
\centering
\begin{tabular*}{\textwidth}{@{\extracolsep{\fill}}lccccccccc@{}}
\toprule
 & $\omega_{de}$ & $\omega_{D}$ & $\omega_{dD}$ & $\omega_{E}$ & $\omega_{eE}$ & $D_D$ & $D_E$ & $D_d$ & $D_{de}$ \\
\midrule
Ref.  & 0.029 & 2.90e-4 & 4.80e-8 & 1.90e-9 & 2.10e-20 & 60.0 & 60.0 & 1.20 & 0.40 \\
PINN  & 0.025 & 1.17e-4 & 1.47e-8 & 4.85e-15 & 1.29e-20 & 1.28e-3 & 8.21e-9 & 0.134 & 0.083 \\
CLIP  & 0.029 & 5.07e-4 & 4.31e-8 & 1.98e-9 & 1.96e-20 & 44.8 & 27.4 & 1.13 & 0.34 \\
\bottomrule
\end{tabular*}
\label{tab:min_results}
\end{table}

\begin{figure}[!htbp]
    \centering
    \includegraphics[width=\linewidth]{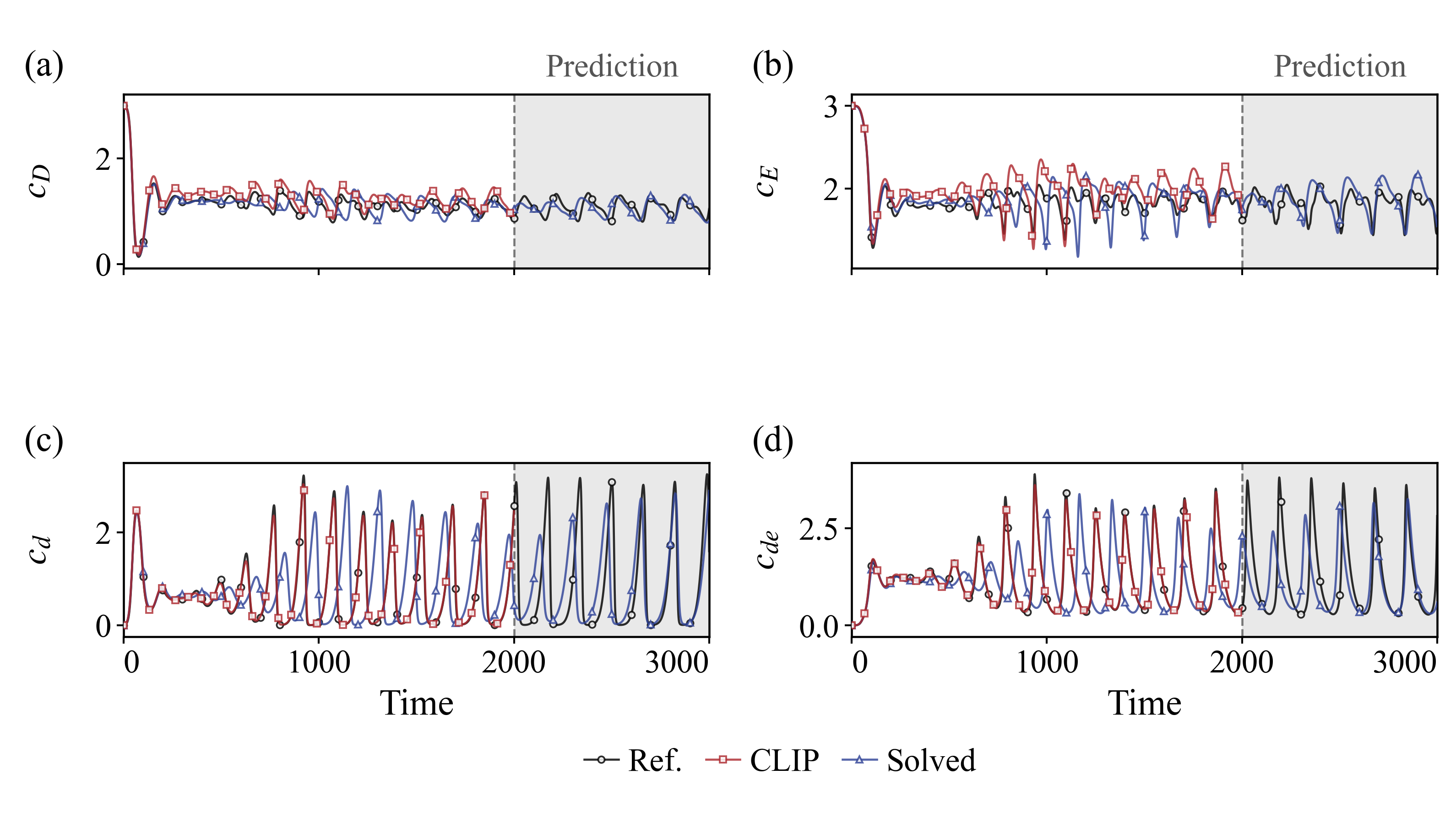}
    \caption{Pointwise comparison of local dynamics in the MinDE system.
    (a,b) Hidden cytosolic states $c_D$ and $c_E$.
    (c,d) Observed membrane states $c_d$ and $c_{de}$.
    Ref., CLIP, and Solved denote the reference, reconstructed, and
    forward-solved trajectories, respectively. Trajectories are sampled
    at the center of each spatial grid, with all concentrations divided
    by $10^6$. The dashed lines mark the end of the reconstruction
    interval at $t=2000$, and the shaded regions indicate prediction over
    $t=2000$--$3000$.}
    \label{fig:min_pointwise_timeseries}
\end{figure}

The identified parameters are further evaluated by numerically solving the MinDE equations from new initial fields.
Figure~\ref{fig:min_reconstruction_combined}(b,d) compares these solutions with those obtained using the reference parameters.
The solutions obtained with the CLIP parameters reproduce the main features of complex pattern formation and evolution under the tested conditions, whereas those obtained with the PINN parameters fail to generate the corresponding structures.
However, differences in local concentrations and pattern locations remain.
The pointwise trajectories in Fig.~\ref{fig:min_pointwise_timeseries} show how the oscillation amplitudes
and phases vary at the sampled locations.
The cytosolic trajectories recover the main concentration levels and oscillatory behavior, but show differences in amplitude and phase.
These results indicate that the identified parameters support complex spatiotemporal dynamics under the tested conditions, while the accuracy of local trajectories remains limited.

To assess the contribution of the training design, Table~\ref{tab:min_ablation} compares the baseline PINN, PINN with curriculum learning but without anchoring, and the full CLIP framework.
The first comparison evaluates the effect of curriculum learning, while the second examines the additional contribution of anchoring.
The ablation results highlight the importance of anchoring in making curriculum learning effective.
Anchoring is expected to be particularly valuable when stage transitions introduce additional degrees of freedom, limiting drift in previously learned estimates during subsequent joint optimization.

\begin{table}[!htbp]
\centering
\caption{Ablation results for parameter identification and hidden-field reconstruction in the Min system under noise-free observations.}
\vspace{6pt}
\label{tab:min_ablation}
\setlength{\tabcolsep}{1pt}
\renewcommand{\arraystretch}{1.05}
\small
\begin{tabular*}{\textwidth}{@{\extracolsep{\fill}}lcc@{}}
\toprule
Method
& Parameter MRAE (\%)
& Hidden-field error (\%) \\
\midrule
PINN         & $67.28$ & $176.30$ \\
PINN + CL    & $65.97$ & $236.81$ \\
\textbf{CLIP} & $30.99 $ & $12.61 $ \\
\bottomrule
\end{tabular*}
\end{table}

Finally, loss landscapes and training trajectories are visualized following Li et al.~\cite{li2018lossvisualizing} to examine the optimization behavior of PINN and CLIP.
Each trajectory is projected onto the plane spanned by its first two principal components, and the loss surface is evaluated on the same plane.
Details are provided in Appendix~\ref{SI:visual}.
Figure~\ref{fig:minde-loss} shows the loss contours and projected trajectories for the baseline PINN and the three CLIP stages.
Each panel uses its own PCA plane and loss scale, and the objectives differ across stages.
The panels therefore describe optimization within each stage rather than provide direct comparisons of curvature.

\begin{figure}[!b]
    \centering
    \includegraphics[width=\linewidth]{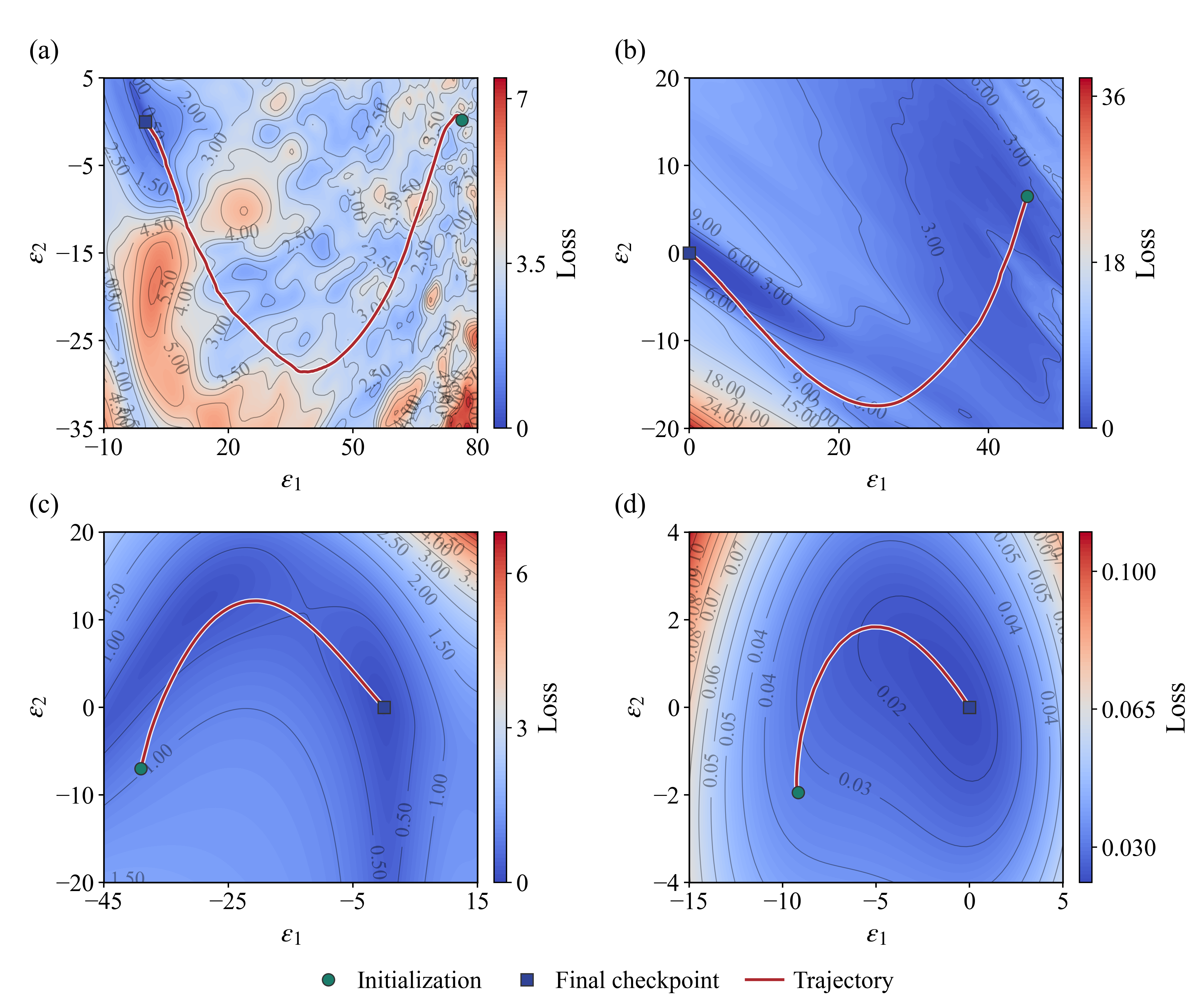}
    \caption{Visualization of the loss landscapes.
    Two-dimensional loss surfaces and projected optimization trajectories
    obtained using PCA. The horizontal and vertical axes denote coordinates
    along the first two PCA directions of the corresponding optimization
    trajectory. Green circles, blue squares, and red curves indicate the
    training initialization, final checkpoint, and optimization trajectory,
    respectively.
    (a) Loss surface for the baseline PINN.
    (b--d) Loss surfaces for the three successive stages of CLIP
    training. Each panel uses its own PCA plane and loss scale.}
    \label{fig:minde-loss}
\end{figure}

For the run and projections shown, CLIP enters the final stage within a region of low loss near its endpoint, and subsequent updates remain largely within this region.
This suggests that the preceding stages provide an initialization close to the final solution, leaving the last stage primarily to refine it.
The PINN trajectory follows a curved path and ends in a region of low loss in the upper left of its projection.
However, this endpoint still corresponds to substantial parameter errors and inaccurate cytosolic reconstructions, as shown in Table~\ref{tab:min_results} and Fig.~\ref{fig:min_reconstruction_combined}(a,c).
Reaching a region of low loss in the projection therefore does not ensure accurate recovery of parameters and hidden states.

Overall, with prescribed initial fields for all species and local membrane observations, the parameters identified by CLIP reproduce the main features of complex pattern formation and evolution from the tested new initial fields.
Differences in local concentrations, oscillation phases, and pattern locations remain.

\section{Summary and Discussion}
In this work, we introduce CLIP, a physics-guided curriculum learning framework for parameter identification and hidden-state reconstruction in partially observed reaction-diffusion systems.
To alleviate the joint optimization difficulties arising from the coupling between physical parameters and hidden states, CLIP initializes the reaction parameters and state representation using approximate local dynamics in reaction-dominated regions.
A proxy constructed from the observed variables selects candidate regions.
Anchored widening transfer is then designed to preserve the learned reaction dynamics while allowing the state representation to adapt to diffusive coupling.
With known governing equations and prescribed initial profiles of all state variables, numerical experiments on three canonical reaction-diffusion systems and a Min-protein system demonstrate improved joint recovery under partial observations and the tested noise conditions.

Across the four systems, the mask increases the fraction of samples with a small diffusion-to-reaction ratio in each hidden component examined.
The selected dynamics, however, are not equally informative about all reaction parameters.
In the $\lambda$--$\omega$ system, for example, the mask mainly selects regions where the amplitude has nearly stabilized, leaving $\alpha$ weakly constrained even though phase evolution remains informative about $\beta$.
Beyond the information contained in the selected regions, unique recovery also depends on the constraints available for the full inverse problem.
The Lotka--Volterra system illustrates a scaling ambiguity when hidden initial profiles and other scale constraints are absent.
Resolving this ambiguity requires additional information beyond the training strategy.

Forward integration with the estimated coefficients provides a second test of the recovered dynamics. In Gray–-Scott, errors in the independently integrated solution are already apparent near the end of the reconstruction interval, even where the network follows the local concentration peaks. In the simulated Min system, the estimated coefficients reproduce the main spatial patterns from new initial fields, with discrepancies in local concentration and oscillation phase.
Since diffusion terms involve second-order spatial derivatives, errors in the reconstructed hidden fields may be partly offset by changes in the diffusion coefficients.

Future work could use limited measurements of the hidden states or additional physical constraints to reduce reliance on complete initial profiles.
Improving observation resolution and sampling could also help resolve hidden spatial structures and improve estimates of the diffusion coefficients.

\section*{Acknowledgments}
This work has been supported by the National Natural Science Foundation of China (Grant Nos. 12588301, 12572247,  and 12432010). Y.Z. also acknowledges the partial support from the Laoshan Laboratory Project under grant No.
LSKJ202202000.

\section*{Author contributions}
Y.Z., Y.C., and H.Z. designed the research. 
H.Z. implemented the algorithm and conducted the numerical experiments. 
Y.Z. managed the project. 
H.Z. wrote the initial manuscript. 
All authors discussed the results, revised the manuscript, and approved the final version.

\section*{Competing interests}
The authors declare no competing interests.


\appendix

\section{Detailed baseline implementations}\label{SI:baseline}
To benchmark parameter identification under partial observations, we compare CLIP against two representative gradient-free baselines: an ensemble Kalman filter (EnKF)-based Bayesian inference approach and the particle swarm optimization (PSO) metaheuristic.
Both methods rely solely on forward PDE simulations and observed data, without requiring gradient information, and thus provide fully non-differentiable references for comparison.

\subsection{Ensemble Kalman filter (EnKF)}
As a representative ensemble-based Bayesian baseline, we employ an EnKF approach for PDE parameter identification under partial observations.
All EnKF experiments are conducted using the open-source DAFI framework \cite{strofer2021dafi}, which is specifically designed for PDE-constrained inverse problems.
In this setting, the unknown physical parameters are treated as part of an augmented state vector and are inferred through an iterative EnKF procedure.

At each iteration, ensemble members are propagated using the same forward PDE solver as in our main experiments.
The simulated states are mapped to the observation space, where all available observations are assimilated without spatial or temporal subsampling.
An ensemble size of 50 members is used throughout all experiments.
The initial ensemble is drawn from a Gaussian prior with unit-scale mean $1.0$ and moderate variance $0.1$.
This configuration provides a fully simulation-driven, derivative-free Bayesian baseline for parameter inference under partial observability.

\subsection{Particle swarm optimization (PSO)}
As another representative gradient-free baseline, we consider particle swarm optimization (PSO) \cite{wang2018pso} for PDE parameter identification.
We adopt a standard global-best PSO formulation with inertia weight, where each particle represents a candidate set of physical parameters.
Particle velocities are updated using fixed acceleration coefficients $C_1 = C_2 = 1.5$ and a constant inertia weight, following the canonical PSO update rule.
A maximum velocity constraint is imposed to avoid unstable updates.

The swarm size is set to 100 particles.
Particle parameters are initialized by random sampling within bounded ranges determined from an initial parameter guess, where the initial guess itself is drawn from the interval $(0,1]$.
Due to the strong nonlinearity of the PDE solver and the presence of partial observations, many randomly sampled parameter sets lead to invalid simulations.
To ensure a feasible initial swarm, particles are repeatedly resampled within the prescribed bounds until 100 valid parameter sets that yield successful PDE simulations are obtained.

The fitness function is defined as the mismatch between simulated and observed dynamics, evaluated only on observable variables and spatial locations.
Each fitness evaluation requires a full forward PDE simulation.
Given the substantial computational cost associated with both initialization and forward solves, the solution obtained after 1{,}000 PSO iterations is reported as the final result.
PSO thus serves as a fully non-differentiable, simulation-driven optimization baseline for comparison.

Unless otherwise stated, all EnKF and PSO hyperparameters follow standard practice and are kept fixed across all test cases, without case-specific tuning.

\section{Detailed ablation experiments}\label{SI:ablation}
This section presents ablation experiments designed to assess the contributions of curriculum learning and anchored widening transfer to parameter identification and hidden-state reconstruction in partially observable reaction–diffusion systems.
We compare three training strategies: a baseline PINN, a baseline PINN with curriculum learning, and a baseline PINN with both curriculum learning and anchored widening transfer, referred to as CLIP.
All three strategies use the same data loss formulation and initial-condition constraints for unobserved variables. 
For the Min example, the parameters are also optimized in log-space, and the PDE residual is evaluated using the same scaling as in the main text, including the characteristic concentration $c_0$.

Baseline PINNs are trained directly on the full spatiotemporal domain without curriculum masking or anchored widening transfer.
Training points are obtained by uniform sampling at a rate of $2\%$ of the full grid.
The total number of optimization steps matches that of CLIP.
To avoid capacity disadvantages, the network architecture is chosen to match the CLIP architecture used in the middle stage, which is the first full-domain PDE training stage.
The optimizer is Adam with a learning rate of $10^{-3}$.
Throughout training, the PDE residual is evaluated using the full RD equations, including all reaction and diffusion terms as well as their associated coefficients.
The PDE residual weight $\eta_\mathrm{pde}$ follows the same growth strategy as in CLIP.

In the second ablation experiment, curriculum learning is introduced without network anchoring. 
The primary stage uses the same network architecture and training configuration as CLIP, including the reaction-dominated sampling mask and the reaction-only PDE residual. 
After the primary stage, training proceeds with the same widened network architecture as the CLIP middle stage, but the anchoring mechanism is disabled. 
All trainable parameters are updated using Adam with a learning rate of $10^{-3}$. 
The second stage training points are sampled uniformly from the full spatiotemporal domain, and training continues until the total number of optimization steps matches that of CLIP. 
The PDE residual weight follows the same update rule as in CLIP during the transition from the primary stage to the full-domain training stage.

The third experiment, CLIP, is described in the main text, and all remaining training details follow the main-text specification.

\begin{table}[!htbp]
\centering
\caption{Ablation results for parameter identification in three
reaction--diffusion systems under noise-free observations.
Values are mean parameter MRAEs (\%).}
\label{tab:canonical_ablation}
\vspace{6pt}
\setlength{\tabcolsep}{5pt}
\renewcommand{\arraystretch}{1.05}
\small
\begin{tabular*}{\textwidth}{@{\extracolsep{\fill}}lccc@{}}
\toprule
Method & $\lambda$--$\omega$ RD & Gray--Scott & Lotka--Volterra \\
\midrule
PINN           & 7.20 & 42.55 & 30.30 \\
PINN + CL      & 5.01 & 33.08 & 31.17 \\
\textbf{CLIP}  & 6.56 & 32.60 & 13.60 \\
\bottomrule
\end{tabular*}
\end{table}

Table~\ref{tab:canonical_ablation} summarizes the ablation results for the three canonical reaction--diffusion systems.
Curriculum learning alone improved parameter identification in the $\lambda$--$\omega$ and Gray--Scott systems, whereas anchoring provided its largest benefit in the Lotka--Volterra system.
These results indicate that the contribution of anchoring depends on the system.

\section{Training and implementation details}\label{SI:hyperparameters}
\subsection{Network architecture and optimization settings}
\label{SI:main_architecture}
The solution field is approximated by a fully connected multi-task physics-informed neural network. 
The network consists of a shared trunk with three hidden layers and one task-specific branch for each state variable, where each branch contains one hidden layer. 
The shared trunk extracts a common spatiotemporal representation from the input coordinates, while the branches output the individual state variables. 
The system-specific widths and activation functions used in the main experiments are summarized in Table~\ref{tab:arch_hyperparams}.
\begin{table}[!htbp]
\centering
\small
\caption{System-specific network architecture and activation functions.}
\label{tab:arch_hyperparams}
\setlength{\tabcolsep}{5pt}
\begin{tabular*}{\textwidth}{@{\extracolsep{\fill}}lcccc@{}}
\toprule
System 
& Trunk width 
& Branch width 
& Widened width 
& Activation \\
\midrule
$\lambda$-$\omega$ 
& 64 
& 64 
& 72 
& $\sin(x)$ \\

Gray-Scott 
& 64 
& 64 
& 72 
& $\sin(x)$ \\

Lotka-Volterra 
& 64 
& 64 
& 72 
& $\sin(x)$ \\

Min 
& 128 
& 128 
& 144 
& $\sin(x)$ \\
\bottomrule
\end{tabular*}
\end{table}

All networks are initialized using Xavier initialization~\cite{glorot2010understanding}, and the unknown physical coefficients are initialized randomly in $[0,1)$. 
Optimization is performed using Adam~\cite{kinga2015adam} following the three-stage CLIP schedule. 

During the diffusion-coupled stage, the network width is increased according to the anchored widening strategy. 
The inherited weights $W_{oo}^{(l)}$ in both trunk and branch layers are anchored across all layers, and the cross blocks $W_{on}^{(l)}$ in the trunk are frozen. 
This preserves the shared representation learned in the reaction-dominated stage while allowing newly introduced degrees of freedom to capture diffusion-induced spatial coupling. 
The stage-wise epoch numbers and learning rates are summarized in Table~\ref{tab:opt_hyperparams}.

\begin{table}[!htbp]
\centering
\small
\caption{System-specific optimization settings used in the three training stages.}
\label{tab:opt_hyperparams}
\setlength{\tabcolsep}{5pt}
\begin{tabular*}{\textwidth}{@{\extracolsep{\fill}}lcccccc@{}}
\toprule
System 
& \multicolumn{3}{c}{Epoch}
& \multicolumn{3}{c}{Learning rate}
\\
\midrule
& Prim. & Mid. & Fine
& Prim. & Mid. (anchored, normal) & Fine
\\
\midrule
$\lambda$-$\omega$ 
& $1000$ & $2000$ & $2000$
& $10^{-3}$ & $10^{-5},\,10^{-3}$ & $10^{-3}$ \\

Gray-Scott 
& $2000$ & $5000$ & $5000$
& $10^{-3}$ & $10^{-8},\,10^{-3}$ & $10^{-4}$ \\

Lotka-Volterra 
& $2000$ & $5000$ & $5000$
& $10^{-3}$ & $10^{-8},\,10^{-3}$ & $10^{-4}$ \\

Min 
& $2000$ & $5000$ & $5000$
& $10^{-3}$ & $10^{-8},\,10^{-3}$ & $10^{-4}$ \\
\bottomrule
\end{tabular*}
\end{table}

The weight of the PDE residual term, $\eta_{\mathrm{pde}}$, is progressively increased during training. 
This strategy prevents the PDE residual from dominating the early optimization before the data loss has reached a stable level. 
The update rule is given in Algorithm~\ref{alg:pde_weight}, and the corresponding parameters for each system and training stage are summarized in Table~\ref{tab:pde_weight_params}. 
The threshold $\zeta_{\mathrm{data}}$ is not applied in the canonical reaction--diffusion cases, whereas for the Min system it is set to $0.01$.
The weights of the initial condition and boundary constraint terms are both set to unity.
Hyperparameters are fixed prior to training and are not tuned separately for individual systems.

\begin{algorithm}[H]
\caption{PDE residual weight $\eta_{\mathrm{pde}}$ update during training.}
\label{alg:pde_weight}
\small
\begin{algorithmic}[1]
\Require Current epoch $e$, data loss $\mathcal{L}_{\mathrm{data}}(e)$, last growth epoch $e_{\mathrm{last}}$
\Require Threshold $\zeta_{\mathrm{data}}$, growth interval $\Delta e$, increment $\Delta \eta$, maximum $\eta_{\max}$
\State $\eta \gets \eta_{\mathrm{pde}}(e)$
\If{$\mathcal{L}_{\mathrm{data}}(e) < \zeta_{\mathrm{data}}$ \textbf{and} $(e - e_{\mathrm{last}}) \ge \Delta e$}
    \State $\eta \gets \min(\eta + \Delta \eta,\ \eta_{\max})$
    \State $e_{\mathrm{last}} \gets e$
\EndIf
\State \Return $\eta$
\end{algorithmic}
\end{algorithm}

\begin{table}[!htbp]
\centering
\small
\caption{Parameters of the adaptive PDE residual weighting strategy (Algorithm~\ref{alg:pde_weight}).}
\setlength{\tabcolsep}{1pt}
\begin{tabular}{lcccccccccccc}
\toprule
 & \multicolumn{3}{c}{$\eta_{\mathrm{init}}$}
 & \multicolumn{3}{c}{$\Delta \eta$}
 & \multicolumn{3}{c}{$\Delta e$}
 & \multicolumn{3}{c}{$\eta_{\max}$}\\
\cmidrule(lr){2-4}
\cmidrule(lr){5-7}
\cmidrule(lr){8-10}
\cmidrule(lr){11-13}
System
 & Prim. & Mid. & Fine
 & Prim. & Mid. & Fine
 & Prim. & Mid. & Fine
 & Prim. & Mid. & Fine \\
\midrule
$\lambda$-$\omega$ RD
 & $1$ & $1$ & $1$
 & $0$ & $0$ & $0$
 & - & - & -
 & - & - & - \\
Gray-Scott
 & $1$ & $10$ & $10$
 & $5\times10^{-3}$ & $5\times10^{-3}$ & $5\times10^{-3}$
 & $1$ & $1$ & $1$
 & $30$ & $30$ & $30$ \\
Lotka-Volterra
 & $1$ & $10$ & $10$
 & $5\times10^{-3}$ & $5\times10^{-3}$ & $5\times10^{-3}$
 & $1$ & $1$ & $1$
 & $30$ & $30$ & $30$\\
Min
 & $10^{-6}$ & $10^{-5}$ & $10^{-5}$
 & $10^{-6}$ & $10^{-6}$ & $10^{-6}$
 & $200$ & $200$ & $200$
 & $10^{-3}$ & $10^{-3}$ & $10^{-3}$ \\
\bottomrule
\end{tabular}
\label{tab:pde_weight_params}
\end{table}

\subsection{Surrogate smoother for reaction-dominated mask construction}\label{SI:hyperparameters-mask}
For noisy observations, the Laplacian used in the reaction-dominated mask is computed from a surrogate smoothed field rather than directly from the raw measurements. 
The surrogate smoother adopts the same network design as the main model, but retains only the branches associated with the observed variables. 
The activation function is chosen consistently with that used in the corresponding main training setting. 
The network is trained by minimizing a Smooth L1 loss over the observation points, using Adam with a learning rate of $10^{-3}$ for 1000 epochs.
The resulting surrogate field is used exclusively for mask construction, while all subsequent physics-informed training is carried out on the original noisy observations.

\section{Visualization of the loss landscape}\label{SI:visual}

To compare the optimization behavior of the baseline PINN and CLIP, we visualize their loss landscapes together with the corresponding optimization trajectories.
Since the parameter space of neural networks is high-dimensional, the loss function is projected onto a two-dimensional subspace, enabling qualitative inspection of the smoothness and curvature of the loss surface encountered during training.

Following the approach of Li et al.~\cite{li2018lossvisualizing}, the high-dimensional loss landscape is projected onto a two-dimensional subspace spanned by the principal directions of the optimization trajectory.
Specifically, we select a reference parameter vector $\theta^\ast$ and two orthogonal directions $\delta$ and $\eta$, and define the projected loss surface as
\begin{equation}
\begin{aligned}
f(\alpha,\beta) = \mathcal{L}(\theta^\ast + \alpha \delta + \beta \eta),
\end{aligned}
\label{eq:loss_landscape}
\end{equation}
where $\mathcal{L}$ denotes the complete PINN loss function and $(\alpha,\beta)$ are scalar coordinates parameterizing the two-dimensional subspace.

The directions $\delta$ and $\eta$ are determined using principal component analysis (PCA) applied to the optimization trajectory.
During training, we collect the model parameters $\{\theta_i\}_{i=0}^{n-1}$ and construct the matrix
\[
M = [\theta_0 - \theta_n;\ \theta_1 - \theta_n;\ \ldots;\ \theta_{n-1} - \theta_n],
\]
from which the first two principal components are extracted and used as the projection directions $\delta$ and $\eta$.
The optimization trajectory is then projected onto this subspace by computing the corresponding $(\alpha,\beta)$ coordinates at each training iteration.

The loss surface is visualized by evaluating $f(\alpha,\beta)$ on a two-dimensional grid chosen to fully cover the projected trajectories of both methods.
This visualization provides qualitative insight into the geometry of the loss landscape and the optimization paths taken by the baseline PINN and CLIP.

\bibliographystyle{aipnum4-1} 
\bibliography{ref}





\end{document}